%% file: draft_arxiv.tex
\documentclass[aps,prb,12pt,showkeys,reprint,longbibliography]{revtex4-1} 
\usepackage{amsmath}
\usepackage{amsthm}
\usepackage{graphicx}
\usepackage[colorlinks]{hyperref}
\usepackage{dcolumn}
\allowdisplaybreaks
\usepackage[usenames,dvipsnames,table]{xcolor}
\usepackage{fancyhdr}
\usepackage{mathptmx}
\usepackage[small,raggedright]{titlesec}
\usepackage{wrapfig}
\usepackage[inline, adjustmargins]{trackchanges}
\addeditor{ari}
\addeditor{jen}

\begin{document}

\thispagestyle{plain}

\title
{
\textcolor{BlueViolet}
{
Effect of dot size on exciton binding energy and electron-hole recombination probability 
in CdSe quantum dots
}
}

\thispagestyle{plain}

\author{Jennifer M. Elward}
\affiliation
{
Department of Chemistry, Syracuse University, Syracuse, New York 13244 USA
}
\author{Arindam Chakraborty}
\affiliation
{
Department of Chemistry, Syracuse University, Syracuse, New York 13244 USA
}
\email[corresponding author:]{archakra@syr.edu}

\date{\today}


\input{sections/sec_abstract_arxiv}

\maketitle

\textcolor{BlueViolet}{\section{Introduction} \label{sec:intro}}
\input{sections/sec_intro}

\textcolor{BlueViolet}{\section{Theory} \label{sec:theory}}
\input{sections/sec_method}
\textcolor{BlueViolet}{\section{Computational Details} \label{sec:comp}}
\input{sections/sec_computation}
\textcolor{BlueViolet}{\section{Results and discussion}\label{sec:results}}
\input{sections/sec_results}
\textcolor{BlueViolet}{\section{Conclusions}\label{sec:conclusions}}
\input{sections/sec_conclusion}


\textcolor{BlueViolet}{\section*{Acknowledgements}}
\label{sec:sec_acknowledgements}
\input{sections/sec_acknowledgements}
\bibliography{ref_version_qd}

\end{document}

%% file: sections/sec_abstract_arxiv.tex
\begin{center}
\begin{abstract}
\noindent\textcolor{BlueViolet}{\rule{14cm}{1.2pt}}
        \colorbox{Apricot}{\parbox{13.7cm}{
\begin{wrapfigure}{r}{2.5in}
\vspace{-10pt}
\includegraphics[width=2.5in]{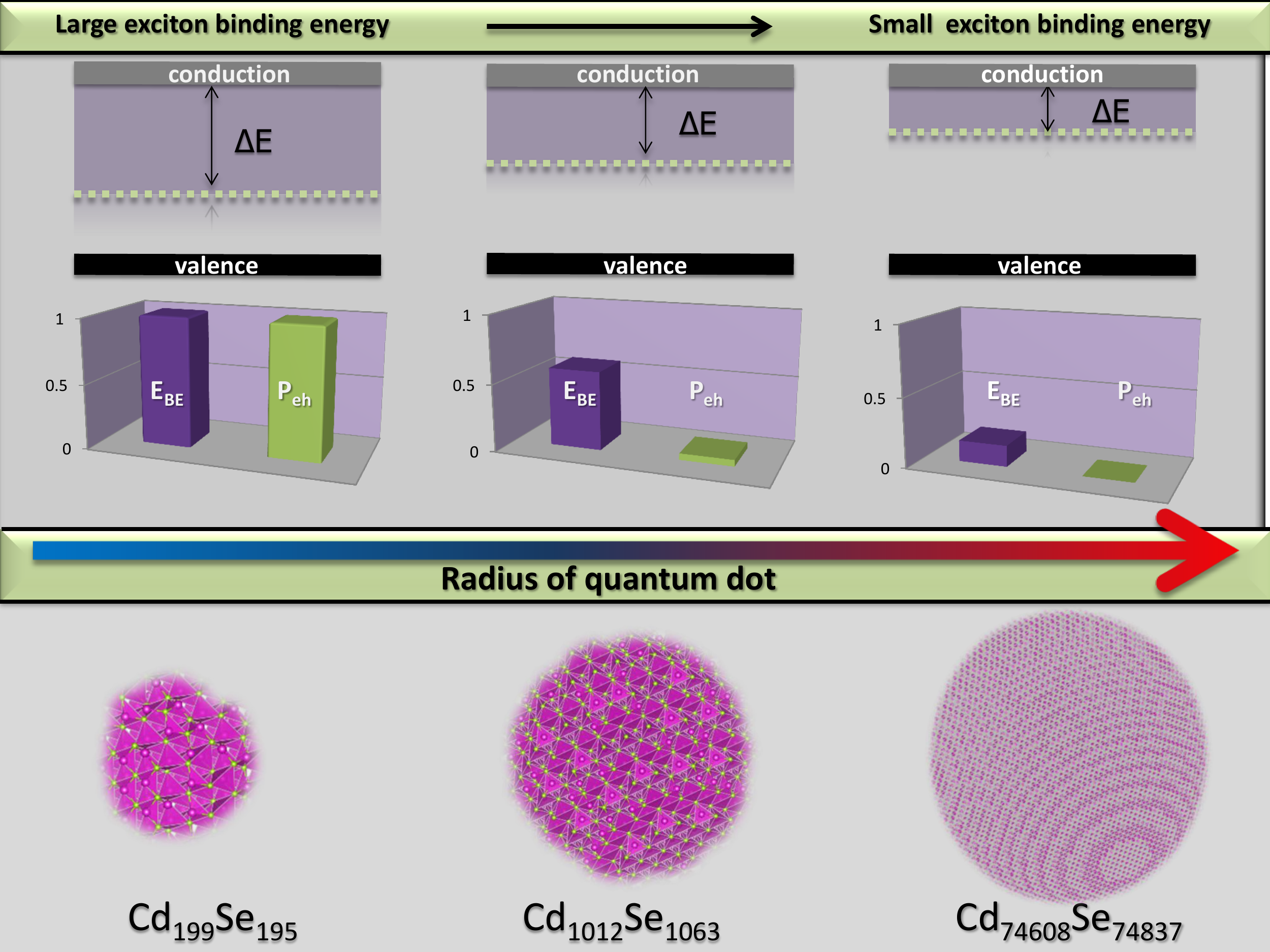}
\vspace{-20pt}
\end{wrapfigure}
Exciton binding energy and electron-hole recombination 
probability are presented as the two important metrics for investigating 
effect of dot size on electron-hole interaction in 
CdSe quantum dots. Direct computation of electron-hole recombination 
probability is challenging because it requires an 
accurate mathematical description of electron-hole wavefunction 
in the neighborhood of the electron-hole coalescence point.
In this work, we address this challenge by solving the electron-hole Schrodinger
equation using the electron-hole explicitly correlated Hartree-Fock (eh-XCHF) method.
The calculations were performed for a series of 
CdSe clusters ranging from 
$\mathrm{Cd}_{20}\mathrm{Se}_{19}$ to $\mathrm{Cd}_{74608}\mathrm{Se}_{74837}$
that correspond to dot diameter range of $1-20$ nm.
The calculated exciton binding energies and electron-hole recombination probabilities 
were found to decrease with increasing dot size.
Both of these quantities were found to scale as  
$D_\mathrm{dot}^{-n}$ with respect to the dot diameter D.
One of the key insights from this study is that 
the electron-hole recombination probability decreases at a much faster rate 
than the exciton binding energy as a function of dot size. 
It was found that an increase in the dot size by a factor of 16.1, resulted in a decrease in the 
exciton binding energy and electron-hole recombination probability by 
a factor of 14.4 and $5.5\times10^{6}$, respectively.
}
}
\noindent\textcolor{BlueViolet}{\rule{14cm}{1.2pt}}
\end{abstract}
\end{center}

%% file: sections/sec_intro.tex
Semiconductor quantum dots and rods have been the focus of intense theoretical and 
experimental research because of inherent size-dependent optical and 
electronic properties. Generation of bound electron-hole pairs (excitons)
and dissociation of excitons into free charge carriers 
are the two important factors that directly impact the 
light-harvesting efficiency of the semiconductor quantum dots. 
The dissociation of excitons is a complex process that is
influenced by various factors such as shape and size 
of the quantum dots,~\cite{zhu2012enhanced,doi:10.1021/ja212032q,doi:10.1021/nn4002825,doi:10.1021/nn200141f,doi:10.1021/nl301291g,doi:10.1021/jp107702b}
presence of surface defects,~\cite{0957-4484-24-13-135703,C3NR00920C,jaeger:064701}
surface ligands,~\cite{doi:10.1021/nn302371q,doi:10.1021/ja9005749} 
and
coupling with phonon modes.~\cite{doi:10.1021/nn800674n,Kelley20101296,Kelley20115254,Kelley2012,Kilina201121641,Hyeon-Deuk20111845,Hyeon-Deuk2012,Kilina2013}
The energetics of the electron-hole interaction in quantum dots is quantified by the exciton dissociation energy 
and has been determined using both theoretical and experimental 
techniques~\cite{musa:243107,doi:10.1021/jp109229u,PhysRevB.63.153304}.
Generation of free charge carrier by exciton dissociated has been 
facilitated by introducing core/shell heterojunctions~\cite{doi:10.1021/ja202752s,doi:10.1021/ja106710m,doi:10.1021/jz4004735},
and applying external and ligand-induced electric fields.~\cite{doi:10.1021/nl300749z,yaacobi2011molecular,PhysRevB.86.045303,doi:10.1021/nl0625022,
blanton:054114,doi:10.1021/nn303719m}
\par
One of the direct routes for enhancing exciton dissociation 
is by modifying the size and shape of quantum dots.
Studies on CdSe and other quantum dots have 
shown that the exciton binding energy decreases with increasing 
dot size.~\cite{PhysRevLett.78.915,
doi:10.1021/nn8006916,
doi:10.1021/nn201681s,PhysRevB.30.2253,ramvall:1104,
PhysRevB.53.9579,
kucur:2333,
CPHC:CPHC200800482,
B508268B}
The size of the quantum dots have significant impact on the 
Auger recombination,~\cite{doi:10.1021/nl901681d,PhysRevLett.91.056404}
 multiple exciton generation~\cite{doi:10.1021/nn2038884,Rabani2010227,doi:10.1021/nn200141f,doi:10.1021/ar3002365},
  and blinking effect in quantum dots~\cite{JBIO:JBIO201000058,doi:10.1021/nn4002825,doi:10.1021/ja212032q}.
In addition to exciton binding energy, 
the spatial distribution of electrons and holes in quantum dots also provides important insight 
into the exciton dissociation process.~\cite{doi:10.1021/jp909118c,doi:10.1021/jp801523m} 
Electron and hole densities $\rho_\mathrm{e}(\mathbf{r})$
and $\rho_\mathrm{h}(\mathbf{r})$ have been widely used to investigate quasi-particle distribution in quantum dots.~\cite{doi:10.1021/ja202752s,doi:10.1021/ja106710m} 
For example in core/shell quantum dots, presence of the heterojunction induces asymmetric spatial 
distribution of electrons and holes which, in turn, facilitates the exciton dissociation. 
Asymmetric electron probability density in the shell region of the core/shell quantum dots has
been attributed to fast electron transfer from the quantum dots.~\cite{doi:10.1021/ja202752s,doi:10.1021/ja106710m,doi:10.1021/jp204420q,doi:10.1021/nn300525b}
\par 
The central challenge in the theoretical investigation of quantum 
dots is efficient computational treatment of
large number of electrons in the system.
For small clusters where all-electron treatment is 
feasible, ground state and excited-state 
calculations have been performed using 
GW Bethe-Salpeter,~\cite{Noguchi2012,LopezDelPuerto2008,DelPuerto2006}
density functional theory (DFT)~\cite{Nguyen201016197,Yang2011405,Albert201115793,Kilin2007342,Liu2009,Chung2009292,Kim2010471}, 
time-dependent DFT
(TDDFT)~\cite{Nadler20131,Abuelela201214674,Fischer2012904,DelBen201116782,Turkowski2009,Li2011157,Yang2012,Yang2013},
and MP2~\cite{Neuhauser20131172}.
For bigger quantum dots where all-electron treatment  is
computationally prohibitive, atomistic semiemperical pseudopotential 
methods have been used extensively.~\cite{PhysRevLett.78.915,PhysRevB.53.9579,PhysRevB.60.1819,PhysRevLett.91.056404,baer:051102}
In this approach, the one-particle Schr{\"o}dinger
equation incorporating the pseudopotential $v_\mathrm{ps}$
\begin{align}
\label{eq:hps}
	\left[
	-\frac{\hbar^2}{2m} \nabla^2 + v_\mathrm{ps}
	\right] \phi_i = \lambda_i \phi_i,
\end{align}
is solved and the eigenfunctions are used in 
construction of the quasiparticle states \cite{PhysRevB.53.9579,PhysRevLett.78.915}. The quasiparticle states 
serve as a basis for both configuration interaction (CI) and perturbation theory 
calculations. Solution of Eq. \eqref{eq:hps} 
is generally obtained by introducing a set of basis functions (typically plane-waves), 
constructing the Hamiltonian matrix in that basis, and diagonalizing it. 
The computational efficiency of CI has been greatly improved by 
using only states near the band gap for construction of the CI space ~\cite{PhysRevB.53.9579,rabani:5355}.  
This technique alleviates the need to compute the entire eigenspectrum of 
the Hamiltonian matrix, however successful implementation of this approach 
requires computation of selected eigenvalues and eigenfunctions of the 
Hamiltonian matrix. Computation of the specific eigenvalues of 
large matrices is challenging and 
various methods such as the 
folded-spectrum method \cite{Wang19942394,Canning200029}, the filter-diagonalization method \cite{Toledo2002256,Neuhauser20131172}, and generalized Davidson method \cite{Vomel20087113,Jordan20124836}
have been specifically developed to address this problem. 
\par  
The main goal of this article is to compare the effect of 
dot size on exciton binding energy and electron-hole recombination 
probability.  
The central quantity of interest for the present work 
is the electron-hole pair density $\rho(\mathbf{r}_\mathrm{e},\mathbf{r}_\mathrm{h})$.
The electron-hole pair density is defined as the probability density of finding an electron 
and a hole in the neighborhood  of
$\mathbf{r}_\mathrm{e}$ and $\mathbf{r}_\mathrm{h}$, respectively. 
The pair density is a mathematically complicated quantity and is
generally obtained from an underlying wavefunction. Direct construction 
of the pair-density is also possible as long as $N-\mathrm{representability}$
can be enforced \cite{mazziotti2007advances}. 
For an interacting electron-hole system, the pair density 
is not equal to the product of electron and hole densities
\begin{align}
	\rho_\mathrm{eh}(\mathbf{r}_\mathrm{e},\mathbf{r}_\mathrm{h})
	\neq
	\rho_\mathrm{e}(\mathbf{r}_\mathrm{e})
	\rho_\mathrm{h}(\mathbf{r}_\mathrm{h}) .
\end{align} 
Furthermore, the electron-hole pair density contains 
information about the correlated spatial distribution of 
the electrons and hole that cannot be obtained from the product of individual electron and hole densities. 
Both electron-hole recombination probability and exciton binding energy can be
computed directly from the pair density. The relationship between the exciton 
binding energy $E_\mathrm{BE}$ and electron-hole pair density is given by the following expression,
\begin{align}
\label{eq:eb}
 E_\mathrm{BE}
 &=
 \int d\mathbf{r}_\mathrm{e} d\mathbf{r}_\mathrm{h} \,\,
 \rho_\mathrm{eh}(\mathbf{r}_\mathrm{e},\mathbf{r}_\mathrm{h})
 r^{-1}_\mathrm{eh} \epsilon^{-1}(\mathbf{r}_\mathrm{e},\mathbf{r}_\mathrm{h}),
\end{align}
where, $\epsilon^{-1}(\mathbf{r}_\mathrm{e},\mathbf{r}_\mathrm{h})$ 
is the inverse dielectric function. 
The electron-hole recombination 
probability, $P_\mathrm{eh}$ is related to the pair density as
\begin{align}
\label{eq:peh}
	P_\mathrm{eh} 
	&=
	\frac{1}{N_\mathrm{e}N_\mathrm{h}}
	\int d\mathbf{r}_\mathrm{e}
	\int_{\mathbf{r}_\mathrm{e}-\frac{\Delta}{2}}^{\mathbf{r}_\mathrm{e}+\frac{\Delta}{2}} 
	d\mathbf{r}_\mathrm{h} \,\,
	\rho_\mathrm{eh}(\mathbf{r}_\mathrm{e},\mathbf{r}_\mathrm{h}),
\end{align}
where $N_\mathrm{e}$ and $N_\mathrm{h}$ are number of electron and holes, 
respectively.
In the above equation, we define electron-hole recombination probability
as the probability of finding a hole 
in a cube of volume $\Delta^3$ centered at the electron position. 
The computation of the recombination probability is especially 
demanding because it requires evaluation of the pair density 
at small interparticle distances. As a consequence, the form of the 
electron-hole wavefunction near the electron-hole coalescence point 
is very important.~\cite{elward:124105,Elward2012182,RefWorks:23,Cancio199011317,Cancio199313246,PhysRevB.54.13575}
 In the present work, we address this challenge by 
using the electron-hole explicitly correlated Hartree-Fock (eh-XCHF) method.~\cite{elward:124105,Elward2012182} 
The eh-XCHF method is 
a variational method where the wavefunction depends explicitly  on the electron-hole interparticle distance and has been used successfully for investigating electron-hole
interaction ~\cite{elward:124105,Elward2012182,blanton:054114}.
\par
The remainder of the article is organized as follows. The theoretical details of the eh-XCHF 
and its computational implementation for CdSe quantum dots are presented 
in \autoref{sec:theory}
and \autoref{sec:comp}, respectively. The results from the 
calculations are presented in \autoref{sec:results}, and the conclusions from the 
study are discussed in \autoref{sec:conclusions}.

%% file: sections/sec_method.tex
In the eh-XCHF method,~\cite{elward:124105,Elward2012182} the electron-hole wavefunction is represented  
by multiplying the mean-field wavefunction with an explicitly correlated function as shown 
in the following equation 
\begin{align}
        \Psi_{\mathrm{eh-XCHF}} = G\Phi^{\mathrm{e}}  \Phi^{\mathrm{h}},
\end{align}
where $\Phi^{\mathrm{e}}$ and $\Phi^{\mathrm{h}}$ are electron and hole Slater determinants and $G$ is a Gaussian-type geminal (GTG) function~\cite{Boys25101960} which is defined as,
\begin{align}
        G(\mathbf{r}^\mathrm{e},\mathbf{r}^\mathrm{h})=\sum_{i=1}^{N_\mathrm{e}}\sum_{j=1}^{N_\mathrm{h}}\sum_{k=1}^{N_\mathrm{g}}
        b_k \mathrm{exp}[-\gamma_k r^2_{ij}].
\end{align}
The GTG function depends on the $r_{\mathrm{eh}}$ term and is responsible for introduction 
of the electron-hole inter-particle distance dependence in 
the eh-XCHF wavefunction. The coefficients $b_k$ and $\gamma_k$ are expansion coefficients which 
are obtained variationally.  
The use of Gaussian-type geminal functions offers three principle advantages.
First, the variational determination of the geminal parameters $\{ b_k,\gamma_k \}$
results in accurate description of the wavefunction near the electron-hole 
coalescence point. This feature is crucial for 
accurate computation of electron-hole recombination probability. 
Second, the integrals of GTG functions with Gaussian-type orbitals (GTO)
can be performed analytically and have been derived earlier by Boys\cite{Boys25101960} and 
Persson et al. \cite{persson:5915} This alleviates the need to approximate the integrals using  
numerical methods. The third advantage of the GTG function is that it allows construction of a compact representation of 
an infinite-order configuration interaction expansion. This can be seen explicitly by 
introduction of the closure relationship,
\begin{align}
	G \vert \Psi_\mathrm{ref} \rangle 
	&=  
	   \underbrace{\sum_{ii'}^{\infty} \vert \Phi^{\mathrm{e}}_i \Phi^{\mathrm{h}}_{i'} \rangle \langle \Phi^{\mathrm{e}}_i \Phi^{\mathrm{h}}_{i'} \vert}
	   _{\mathbf{1}}
	   G \vert \Psi_\mathrm{ref} \rangle.
\end{align}
The electron-hole interaction was described using the effective electron-hole Hamiltonian\cite{PhysRevB.42.1713,PhysRevB.54.13575,
PhysRevLett.87.186402,PhysRevB.73.165305,woggon1997springer,Brasken2001775,Corni2003436,Corni2003853141,Corni2003453131,Vanska20064035,Vanska2010,Sundholm201296} which is defined in the following equation 
\begin{align}
\label{eq:heh}
	H &= 
	\sum_{ij} \langle i \vert 
	\frac{-\hbar^2}{2m_\mathrm{e}} + v_\mathrm{ext}^\mathrm{e}
	\vert j\rangle e_i^\dagger e_j \\ \notag
    &+ \sum_{ij} \langle i \vert 
    \frac{-\hbar^2}{2m_\mathrm{h}}+ v_\mathrm{ext}^\mathrm{h}
    \vert j\rangle     h^\dagger_i h_j \\ \notag
    &+ \sum_{iji'j'}    \langle iji'j' \vert
    \epsilon^{-1} r_\mathrm{eh}^{-1}
    \vert iji'j' \rangle
    e^\dagger_i e_j h_{i'}^\dagger h_{j'} \\ \notag
    &+\sum_{ijkl} w_{ijkl}^\mathrm{ee} e^\dagger_i e_j^\dagger e_l e_k 
    +\sum_{ijkl} w_{ijkl}^\mathrm{hh} h^\dagger_i h_j^\dagger h_l h_k.
\end{align}
The 
effective electron-hole Hamiltonian provides a computationally efficient
route for investigating large systems and in the present work was
used for investigating CdSe clusters in the range of $\mathrm{Cd}_{20}\mathrm{Se}_{19}$ to $\mathrm{Cd}_{74608}\mathrm{Se}_{74837}$. 
We have also developed eh-XCHF method using a pseudopotential~\cite{elward_atomistic},
but the current implementation is restricted to cluster sizes of 200 atoms
and cannot be applied to large dot sizes. 
\par
The effective Hamiltonian in Eq. \eqref{eq:heh} was used in combination with parabolic potential which 
has been used extensively~\cite{RefWorks:44,RefWorks:43,RefWorks:45,RefWorks:46,RefWorks:49,RefWorks:48,RefWorks:47,Karimi20114423,Nammas20114671,Rezaei20114596} for approximating the confining potential in quantum dots and wires.
The electron and hole external potentials $v^\alpha_\mathrm{ext}$ were expressed as 
\begin{align}
	v^\alpha_\mathrm{ext} = \frac{1}{2} k_\alpha \vert \mathbf{r}_\alpha \vert^2 \quad \alpha = \mathrm{e,h}.
\end{align}
The form of the external potential directly impacts the
electron-hole pair density and is important for 
accurate computation of the binding energy and recombination 
probability. In this work, we have developed a particle number based search
procedure for determining the external 
potential.  The central idea of this method is to find an external potential 
such that the computed 1-particle electron and hole densities are spatially
confined within the volume of the quantum dot. Mathematically, this is 
implemented by obtaining the force constant $k$ by the 
following minimization process
\begin{align}
	\min_{k_\alpha^\mathrm{min}}
	\left(
	N_\alpha - \int_0^{\frac{D_\mathrm{dot}}{2}} dr r^2 \int d\Omega  \rho_\alpha(\mathbf{r})		[v^\alpha_\mathrm{ext}]
	\right)^2, 
\end{align}
where $\alpha = \mathrm{e,h}$, $d\Omega=sin\theta d\theta d\phi$,
$D_\mathrm{dot}$ is the dot diameter,
 and $k_\alpha^\mathrm{min}$ is the smallest force constant 
that satisfies the above minimization conditions. The single-particle density is a functional of the external 
potential and is denoted explicitly in the above equation. 
\par
The eh-XCHF wavefunction is obtained variationally
by minimizing the eh-XCHF energy 
\begin{align}
	E_\mathrm{eh-XCHF} &= \min_{G,\Phi^\mathrm{e},\Phi^\mathrm{h}}\frac{\langle \Psi_\mathrm{eh-XCHF}\vert H \vert \Psi_\mathrm{eh-XCHF}\rangle}
	                        {\langle \Psi_\mathrm{eh-XCHF} \vert \Psi_\mathrm{eh-XCHF}\rangle}.
\end{align}
Instead of evaluating the above equation directly, it is more efficient 
to first transform the operators and then perform the integration over the
coordinates. The transformed operators are obtained by performing
congruent transformation~\cite{PhysRevA.86.062504,bayne_pios} 
which is defined as follows
\begin{align}
	\tilde{H} = G^\dagger H G \\
	\tilde{1} = G^\dagger H G.
\end{align}
The eh-XCHF energy is obtained from the transformed operators using
the following expression
\begin{align}
	E_\mathrm{eh-XCHF} &= \frac{\langle \Phi^\mathrm{e},\Phi^\mathrm{h}\vert \tilde{H}  \vert \Phi^\mathrm{e},\Phi^\mathrm{h}\rangle}
	 {\langle \Phi^\mathrm{e},\Phi^\mathrm{h} \vert \tilde{1} \vert
	                       \Phi^\mathrm{e},\Phi^\mathrm{h}\rangle}.
\end{align}
The above equation allows us to reduce the minimization over the 
electron and hole Slater determinants in terms of coupled 
self-consistent field (SCF) equations as shown below \cite{doi:10.1021/jp0634297}
\begin{align}
	\mathbf{F}_G^\mathrm{e}[\mathbf{C}^\mathrm{h}]  \mathbf{C}^\mathrm{e}
	&=
	\lambda^\mathrm{e} \mathbf{S}_G^\mathrm{e}  \mathbf{C}^\mathrm{e} \\
	\mathbf{F}_G^\mathrm{h}[\mathbf{C}^\mathrm{e}]  \mathbf{C}^\mathrm{h}
	&=
	\lambda^\mathrm{h} \mathbf{S}_G^\mathrm{h}  \mathbf{C}^\mathrm{h} .
\end{align}
This is identical to the Roothaan-Hall equation
where $\mathbf{F}_G^\mathrm{e}$ and $\mathbf{F}_G^\mathrm{e}$
are Fock matrices for electron and holes, respectively. The subscript $G$
in the above expression denotes that the Fock operators were obtained 
from the congruent transformed Hamiltonian and contains contribution 
from the geminal operator. The functional form of the congruent transformed
operators and the Fock operators have been derived earlier and can be 
found in Ref.~\citenum{Elward2012182}.
The single-particle basis for electrons and holes are constructed from the 
eigenfunctions of zeroth order single-particle Hamiltonian
\begin{align}
	H_0^\alpha \phi_i^\alpha = E_i^\alpha \phi_i^\alpha \quad \alpha = \mathrm{e,h}. 
\end{align}
where the zeroth-order Hamiltonian is obtained from $H$ using the following 
limiting condition
\begin{align}
\label{eq:ho}
	H_0^\mathrm{e} + H_0^\mathrm{h} = \lim_\mathrm{{r_{eh} \rightarrow \infty}} H.
\end{align}
\par
Equations \eqref{eq:heh} to \eqref{eq:ho} summarize the key steps of the 
eh-XCHF method.


%% file: sections/sec_computation.tex
The material parameters for the CdSe quantum dots used in the 
electron-hole Hamiltonian in Eq. \eqref{eq:heh} were 
obtained from Ref.~\citenum{PhysRevB.73.165305}  is presented in \autoref{tab:cdse_param}.
\begin{table} 
 \begin{center} 
  \caption{\textbf{Material parameters for the CdSe quantum dots used in the electron-hole Hamiltonian}}
   \label{tab:cdse_param} 
    \begin{tabular}{ c c } 
    \hline
     Property  &      Value (Atomic units)~\cite{PhysRevB.73.165305}  \\ 
     \hline 
      $m_\mathrm{e}$         & $0.13$     \\
      $m_\mathrm{h}$         & $0.38$     \\
      $\epsilon$             & $6.2$     \\
     \hline 
    \end{tabular}
  \end{center} 
\end{table} 
The single-particle basis was constructed using a set of ten
s,p,d GTOs and the details of the basis functions and the 
external potential parameters used in the calculations are presented in  \autoref{tab:param_gto}. 
\begin{table*}
  \begin{center}
   \caption{\textbf{Parameters for the external potential and the GTOs used in the eh-XCHF calculation. All values are given in atomic units.}}
   \label{tab:param_gto}
    \begin{tabular}{c c c c c }
    \hline
       $\mathrm{D_{dot}}$(nm) & $k_\mathrm{e}$   &  $k_\mathrm{h}$ & $\alpha_\mathrm{e}$ & $\alpha_\mathrm{h}$  \\
      \hline
      1.19                &  $2.66\times10^{-2}$ & $9.10\times10^{-3}$  & $2.94\times10^{-2}$ & $2.94\times10^{-2}$ \\   
      1.69                &  $6.22\times10^{-3}$ & $2.13\times10^{-3}$  & $1.42\times10^{-2}$ & $1.42\times10^{-2}$ \\   
      2.71                &  $1.10\times10^{-3}$ & $3.76\times10^{-4}$  & $5.98\times10^{-3}$ & $5.98\times10^{-3}$ \\   
      2.96                &  $8.10\times10^{-4}$ & $2.77\times10^{-4}$  & $5.13\times10^{-3}$ & $5.13\times10^{-3}$ \\
      3.23                &  $5.52\times10^{-4}$ & $1.89\times10^{-4}$  & $4.24\times10^{-3}$ & $4.24\times10^{-3}$ \\
      3.76                &  $3.09\times10^{-4}$ & $1.06\times10^{-4}$  & $3.17\times10^{-3}$ & $3.17\times10^{-3}$ \\
      4.79                &  $1.20\times10^{-4}$ & $4.12\times10^{-4}$  & $1.98\times10^{-3}$ & $1.98\times10^{-3}$ \\ 
      6.58                &  $3.38\times10^{-5}$ & $1.16\times10^{-5}$  & $1.05\times10^{-3}$ & $1.05\times10^{-3}$ \\ 
      9.98                &  $6.41\times10^{-6}$ & $2.19\times10^{-6}$  & $4.57\times10^{-4}$ & $4.57\times10^{-4}$ \\
      15.0                &  $1.26\times10^{-6}$ & $4.33\times10^{-7}$  & $2.03\times10^{-4}$ & $2.03\times10^{-4}$ \\ 
      19.9                &  $4.01\times10^{-7}$ & $1.37\times10^{-7}$  & $1.14\times10^{-4}$ & $1.14\times10^{-4}$ \\
     \hline 
    \end{tabular}
  \end{center}  
\end{table*}   
A set of three geminal functions were used for each dot size, where the geminal parameters were optimized 
variationally. The optimized parameters for all the dot sizes are 
presented in \autoref{tab:param_gem}.
\begin{table*}
  \begin{center}
   \caption{\textbf{Optimized geminal parameters obtained by minimizing the eh-XCHF energy.
	The first set of geminal parameters were set to $b_1=1$ and $g_1=0$ and the details are
	presented in the text.
    All values are given in atomic units.}}
   \label{tab:param_gem}
    \begin{tabular}{c c c c c}
    \hline
       $\mathrm{D_{dot}}$(nm) & $b_2$   &  $b_3$ & $\gamma_2$ & $\gamma_3$  \\
      \hline
      1.19     & $3.35\times10^{-2}$ & $1.20\times10^{-2}$ & $1.08\times10^{-1}$ & $4.59\times10^{-2}$ \\
      1.69     & $6.21\times10^{-2}$ & $-1.22\times10^{-2}$ & $3.21\times10^{-1}$ & $1.23\times10^{-1}$ \\
      2.71     & $3.10\times10^{-2}$ & $1.07\times10^{-2}$ & $3.22\times10^{-2}$ & $1.48\times10^{-2}$ \\
      2.96     & $2.88\times10^{-2}$ & $1.07\times10^{-2}$ & $3.19\times10^{-2}$ & $1.47\times10^{-2}$ \\
      3.23     & $2.34\times10^{-2}$ & $4.31\times10^{-3}$ & $1.96\times10^{-2}$ & $1.22\times10^{-2}$ \\         
      3.76     & $2.87\times10^{-2}$ & $1.06\times10^{-2}$ & $2.09\times10^{-2}$ & $1.37\times10^{-2}$ \\
      4.79     & $2.03\times10^{-2}$ & $7.41\times10^{-3}$ & $1.74\times10^{-2}$ & $1.24\times10^{-2}$ \\
      6.58     & $2.55\times10^{-2}$ & $1.06\times10^{-2}$ & $1.86\times10^{-2}$ & $1.37\times10^{-2}$ \\
      9.98     & $4.96\times10^{-2}$ & $1.38\times10^{-2}$ & $1.89\times10^{-2}$ & $1.49\times10^{-2}$ \\
      15.0     & $4.85\times10^{-2}$ & $9.42\times10^{-3}$ & $1.79\times10^{-2}$ & $1.36\times10^{-2}$ \\
      19.9     & $1.69\times10^{-2}$ & $7.41\times10^{-3}$ & $1.60\times10^{-2}$ & $1.24\times10^{-2}$ \\
     \hline 
    \end{tabular}
  \end{center}  
\end{table*}   
The first set of geminal parameters were always set to $b_1=1$ and $\gamma_1=0$
to ensures that  the eh-XCHF energy is always bounded 
from above by the mean-field energy during the geminal optimization.~\cite{elward:124105,Elward2012182} 

%% file: sections/sec_results.tex
\subsection{Exciton binding energy}
\begin{figure}[ht] 
\begin{center} 
\includegraphics[width=100mm]{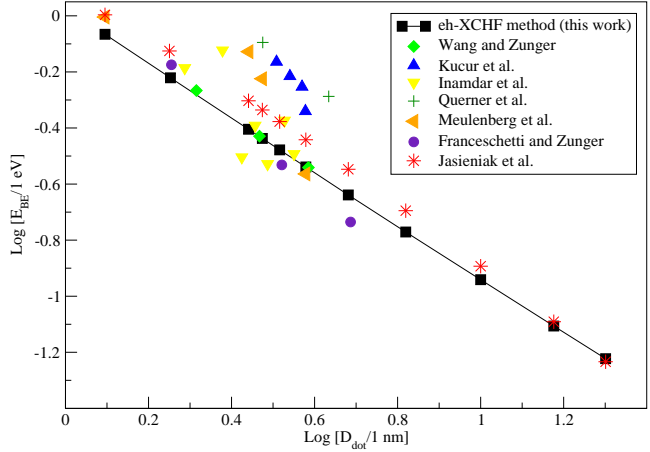}
\caption{Log of binding energy ($E_{BE}$) versus log of diameter for CdSe quantum dots. 
The values from the eh-XCHF calculations are compared with 
results from earlier studies by
Wang et al.~\cite{PhysRevB.53.9579},
Franceschetti et al.~\cite{doi:10.1021/nn8006916},
Meulenberg et al.~\cite{PhysRevLett.78.915},
Jasieniak et al.~\cite{doi:10.1021/nn201681s},
Kucur et al.~\cite{kucur:2333},
Inamdar et al.~\cite{CPHC:CPHC200800482}, and
Querner et al.~\cite{B508268B}
The details of the comparison are presented in the text.
}
\label{fig:log_log_be}
\end{center}
\end{figure}

The exciton binding energy was computed for a series of CdSe clusters ranging from 
$\mathrm{Cd}_{20}\mathrm{Se}_{19}$ to $\mathrm{Cd}_{74608}\mathrm{Se}_{74837}$.
The approximate diameters of these quantum dots are in the range of  
$1$ to $20\mathrm{nm}$, respectively and the results are presented in 
\autoref{tab:cluster_tab}.
\begin{table}
  \begin{center}
   \caption{\textbf{Exciton binding energy calculated using eh-XCHF method as function of dot diameter.}}
   \label{tab:cluster_tab}
    \begin{tabular}{c l l}
    \hline
       $\mathrm{D_{dot}}$(nm) & $\mathrm{Cd}_{x}\mathrm{Se}_{y}$   &  $E_\mathrm{BE}$(eV)  \\
      \hline
      1.19                & $\mathrm{Cd}_{20}\mathrm{Se}_{19}$               & 0.859 \\
      1.69                & $\mathrm{Cd}_{47}\mathrm{Se}_{57}$               & 0.601 \\
      2.71                & $\mathrm{Cd}_{199}\mathrm{Se}_{195}$             & 0.394 \\ 
      2.96                & $\mathrm{Cd}_{232}\mathrm{Se}_{257}$             & 0.365 \\ 
      3.23                & $\mathrm{Cd}_{311}\mathrm{Se}_{352}$             & 0.333  \\
      3.76                & $\mathrm{Cd}_{513}\mathrm{Se}_{515}$             & 0.289  \\
      4.79                & $\mathrm{Cd}_{1012}\mathrm{Se}_{1063}$           & 0.230 \\ 
      6.58                & $\mathrm{Cd}_{2704}\mathrm{Se}_{2661}$           & 0.170 \\ 
      9.98                & $\mathrm{Cd}_{9338}\mathrm{Se}_{9363}$           & 0.115 \\ 
      15.0                & $\mathrm{Cd}_{31534}\mathrm{Se}_{31509}$         & 0.078 \\ 
      19.9                & $\mathrm{Cd}_{74608}\mathrm{Se}_{74837}$         & 0.060 \\
     \hline 
    \end{tabular}
  \end{center}  
\end{table}   
It is seen that binding energy decreases as the size of the quantum dot increases.
This trend is in agreement with earlier results~\cite{PhysRevLett.78.915,doi:10.1021/nn8006916,doi:10.1021/nn201681s}.
In \autoref{fig:log_log_be}, the computed binding energies are compared with previously reported experimental and theoretical results
~\cite{PhysRevLett.78.915,doi:10.1021/nn8006916,doi:10.1021/nn201681s,PhysRevB.53.9579,kucur:2333,CPHC:CPHC200800482,B508268B}.
For $D_\mathrm{dot}$ equal to $1.8$, $3.32$ and $4.82\,\mathrm{nm}$, Franceschetti and  
Zunger have computed binding energies using atomistic pseudopotential based 
configuration interaction method~\cite{PhysRevLett.78.915} and the exciton binding 
energies shown in \autoref{fig:log_log_be} were obtained from 
the tabulated values in Ref.~\citenum{PhysRevLett.78.915}. 
In a recent combined experimental and theoretical investigation, Jasieniak et al.~\cite{doi:10.1021/nn201681s} have reported size-dependent valence and conduction band energies of CdSe quantum dots. The values from the Jasieniak et al. studies in \autoref{fig:log_log_be} were obtained from the least-square fit equation provided in Ref.~\citenum{doi:10.1021/nn201681s}.
The remaining data points were obtained from the plot in Ref.~\citenum{doi:10.1021/nn201681s}.
The log-log plot in \autoref{fig:log_log_be} shows that the computed binding energy 
is described very well by a linear-fit and the exciton binding energy scales as $D^{-n}$ with
respect to the dot size.  This observation is consistent with trend observed in earlier studies.~\cite{PhysRevLett.78.915,doi:10.1021/nn8006916,doi:10.1021/nn201681s}
We find that the exciton binding energy from the eh-XCHF calculations 
are in very good agreement with the 
atomistic pseudopotential calculations by Wang et al.~\cite{PhysRevB.53.9579} and 
Franceschetti et al.~\cite{PhysRevLett.78.915} Comparing between eh-XCHF and 
Jasieniak et al.~\cite{doi:10.1021/nn201681s} results show that the eh-XCHF values are
lower than the Jasieniak et al. values for small dot sizes, but the difference becomes
smaller with increasing dot size. One possible explanations for this observation is
that the smaller quantum dots have high surface to volume ratios and their optical
properties are dominated by surface effects~\cite{jasieniak2007cd,luther2013stoichiometry}
that are not currently included in the eh-XCHF calculations. The plot in \autoref{fig:log_log_be}
highlights the ability of the eh-XCHF method to predict exciton binding energies for large quantum dots.
\subsection{Electron-hole Coulomb energy}
\begin{figure}[ht] 
\begin{center} 
\includegraphics[width=100mm]{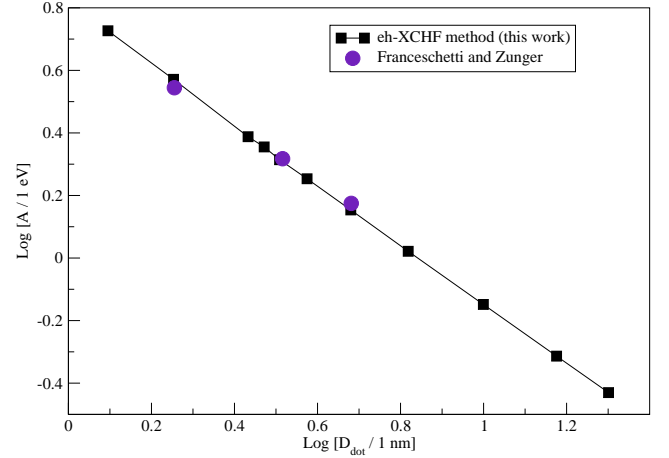}
\caption{Log of Coulomb energy ($A$) for CdSe quantum dots versus log of diameter of quantum dot. }
\label{fig:log_log_coul}
\end{center}
\end{figure}

Another important quantity that is directly related to the
electron-hole interaction is the electron-hole Coulomb energy. We have used the definition given by 
Franceschetti and  Zunger~\cite{PhysRevLett.78.915} and calculated
the electron-hole Coulomb energy  using the following expression 
\begin{align}
\label{eq:coul}
 A
 &=
 \int d\mathbf{r}_\mathrm{e} d\mathbf{r}_\mathrm{h} \,\,
 \rho_\mathrm{eh}(\mathbf{r}_\mathrm{e},\mathbf{r}_\mathrm{h})
 r^{-1}_\mathrm{eh} .
\end{align}
In \autoref{fig:log_log_coul}, we have compared the electron-hole Coulomb energy with the pseudopotential+CI calculations
by Franceschetti and  Zunger and the results were found to be in good agreement with each other. 
If the dielectric function is approximated by a constant, then the exciton binding energy 
is related to Coulomb energy by the expression
\begin{align}
	E_\mathrm{BE} = \epsilon^{-1} A 
	\quad \mathrm{for} \quad \epsilon(\mathbf{r}^\mathrm{e},\mathbf{r}^\mathrm{h}) = \epsilon.
\end{align}
The Coulomb energy is a very important quantity because it allows us to 
directly compare the quality of electron-hole pair density 
without introducing any additional approximation due to the 
choice of the  dielectric function used for computation of 
the binding energy. The good agreement between the two methods provides
important verification of the implementation of the eh-XCHF method. 
\subsection{Recombination probability}
\begin{figure}[ht] 
\begin{center} 
\includegraphics[width=100mm]{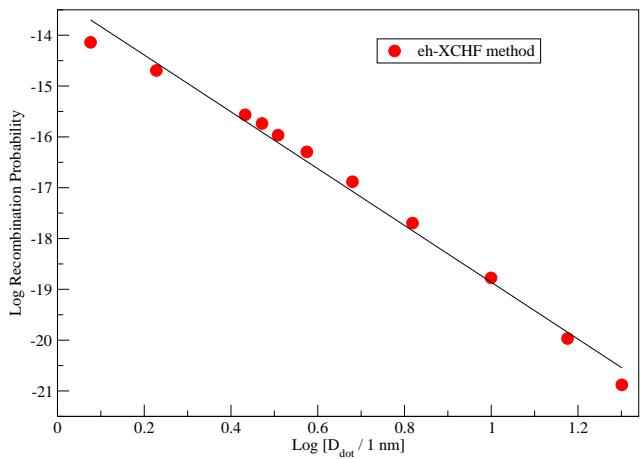}
\caption{Log of recombination probability ($\mathrm{P_{eh}}$) of CdSe quantum dots versus log of diameter of quantum dot. }
\label{fig:log_log_peh}
\end{center}
\end{figure}

\begin{figure}[ht] 
\begin{center} 
\includegraphics[width=100mm]{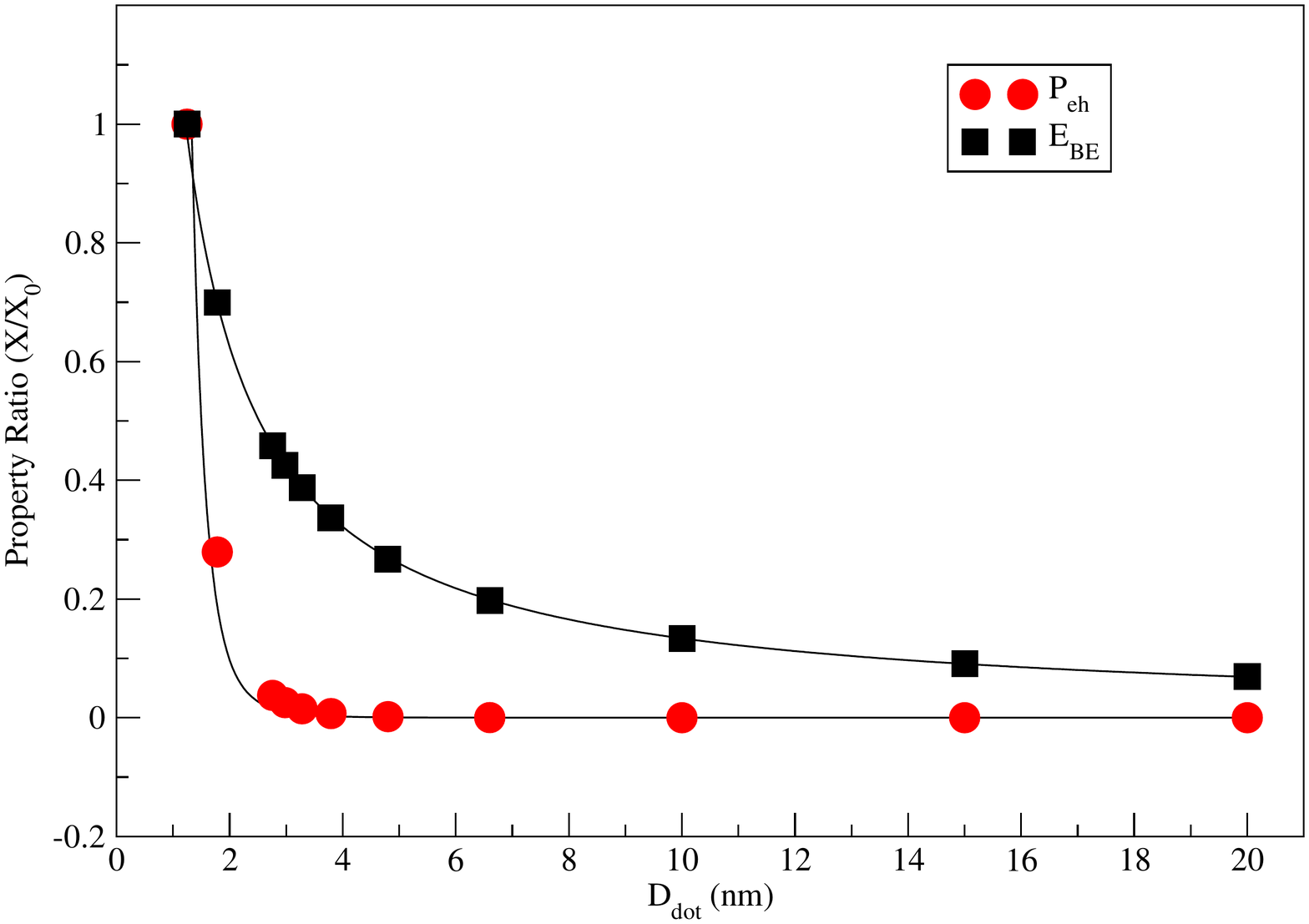}
\caption{Comparison of $E_\mathrm{BE}$ and $P_\mathrm{eh}$ relative properties versus $r_\mathrm{dot}$. }
\label{fig:fig_comp_prop}
\end{center}
\end{figure}

In addition to exciton binding energies, electron-hole recombination probabilities were also calculated. 
Using the expression in Eq. \eqref{eq:peh}, the electron-hole pair density from the eh-XCHF method was 
used in the computation of electron-hole recombination probabilities and the 
results are presented in \autoref{fig:log_log_peh}.
 A log-log plot of $P_\mathrm{eh}$ versus 
$D_\mathrm{dot}$ indicates that the recombination probability also follows $D_\mathrm{dot}^{-n}$
dependence with dot diameter. One of the key results from this study is that the
electron-hole recombination probability decreases at a much faster rate that the exciton-binding energy with increasing dot
size. This is illustrated in \autoref{fig:fig_comp_prop},
where comparison of the relative binding energy and recombination probability 
is presented with respect to dot size. It was found that for a factor of 16.1 change in the dot diameter, the exciton binding energy
and the recombination probability decrease by a factor of $14.4$ and $5.5 \times 10^{6}$, respectively. 
The linear regression equations of the Coulomb energy, exciton binding energy and electron-hole 
recombination probability as function of dot diameter are summarized in \autoref{tab:func_tab}.
\begin{table}
 \begin{center} 
  \caption{\textbf{Linear regression equation of 
  Coulomb energy, exciton binding energy, and electron-hole 
recombination probability with respect to dot diameter}}
   \label{tab:func_tab} 
    \begin{tabular}{ l c } 
    \hline
     Property  &      Equation \\ 
     \hline 
      $\mathrm{log}[\mathrm{A} / \mathrm{eV}]$              & $-0.958\,\mathrm{log[D / \mathrm{nm}}] + 0.8106$     \\
      $\mathrm{log}[\mathrm{E}_\mathrm{{BE}} / \mathrm{eV}]$         & $-0.958\,\mathrm{log [D / \mathrm{nm}}] + 0.0182$     \\
      $\mathrm{log} \mathrm{P}_\mathrm{{eh}}$         &     $ -5.590\,\mathrm{log [D / nm}] - 13.271$     \\
     \hline 
    \end{tabular}
  \end{center} 
\end{table} 
It is seen that the slope for the recombination is substantially higher than the binding energy.

%% file: sections/sec_conclusion.tex
In conclusion, we have presented a multifaceted  
study of effect of dot size on electron-hole 
interaction in CdSe quantum dots. The electron-hole 
explicitly correlated Hartree-Fock method  was used for 
computation of exciton binding energy and electron-hole 
recombination probability. 
It was found that both exciton binding energy and 
electron-hole recombination probability decreases with 
increasing dot size and both quantities 
scale as $D_\mathrm{dot}^{-n}$ with respect to the
diameter of the quantum dot.
The computed exciton binding energies 
were found to be in 
good agreement with previously reported results.
 One of significant results 
from these calculations is that that the electron-hole 
recombination probability decreases at a 
substantially higher rate than the binding energy with 
increasing dot size. Changing the dot size by a factor of 14
resulted in a decrease in the electron-hole recombination 
probability by a factor of $10^6$. We believe this to be a 
significant result that can enhance our understanding 
of the electron-hole interaction in quantum dots.

%% file: sections/sec_acknowledgements.tex
We gratefully acknowledge the support from Syracuse
University for this work.

%% file: draft_arxiv.bbl
\begin{thebibliography}{113}%
\makeatletter
\providecommand \@ifxundefined [1]{%
 \@ifx{#1\undefined}
}%
\providecommand \@ifnum [1]{%
 \ifnum #1\expandafter \@firstoftwo
 \else \expandafter \@secondoftwo
 \fi
}%
\providecommand \@ifx [1]{%
 \ifx #1\expandafter \@firstoftwo
 \else \expandafter \@secondoftwo
 \fi
}%
\providecommand \natexlab [1]{#1}%
\providecommand \enquote  [1]{``#1''}%
\providecommand \bibnamefont  [1]{#1}%
\providecommand \bibfnamefont [1]{#1}%
\providecommand \citenamefont [1]{#1}%
\providecommand \href@noop [0]{\@secondoftwo}%
\providecommand \href [0]{\begingroup \@sanitize@url \@href}%
\providecommand \@href[1]{\@@startlink{#1}\@@href}%
\providecommand \@@href[1]{\endgroup#1\@@endlink}%
\providecommand \@sanitize@url [0]{\catcode `\\12\catcode `\$12\catcode
  `\&12\catcode `\#12\catcode `\^12\catcode `\_12\catcode `\%12\relax}%
\providecommand \@@startlink[1]{}%
\providecommand \@@endlink[0]{}%
\providecommand \url  [0]{\begingroup\@sanitize@url \@url }%
\providecommand \@url [1]{\endgroup\@href {#1}{\urlprefix }}%
\providecommand \urlprefix  [0]{URL }%
\providecommand \Eprint [0]{\href }%
\providecommand \doibase [0]{http://dx.doi.org/}%
\providecommand \selectlanguage [0]{\@gobble}%
\providecommand \bibinfo  [0]{\@secondoftwo}%
\providecommand \bibfield  [0]{\@secondoftwo}%
\providecommand \translation [1]{[#1]}%
\providecommand \BibitemOpen [0]{}%
\providecommand \bibitemStop [0]{}%
\providecommand \bibitemNoStop [0]{.\EOS\space}%
\providecommand \EOS [0]{\spacefactor3000\relax}%
\providecommand \BibitemShut  [1]{\csname bibitem#1\endcsname}%
\let\auto@bib@innerbib\@empty
\bibitem [{\citenamefont {Zhu}\ and\ \citenamefont
  {Lian}(2012)}]{zhu2012enhanced}%
  \BibitemOpen
  \bibfield  {author} {\bibinfo {author} {\bibfnamefont {Haiming}\ \bibnamefont
  {Zhu}}\ and\ \bibinfo {author} {\bibfnamefont {Tianquan}\ \bibnamefont
  {Lian}},\ }\bibfield  {title} {\enquote {\bibinfo {title} {Enhanced multiple
  exciton dissociation from cdse quantum rods: The effect of nanocrystal
  shape},}\ }\href {\doibase 10.1021/ja304724u} {\bibfield  {journal} {\bibinfo
   {journal} {Journal of the American Chemical Society}\ }\textbf {\bibinfo
  {volume} {134}},\ \bibinfo {pages} {11289--11297} (\bibinfo {year}
  {2012})}\BibitemShut {NoStop}%
\bibitem [{\citenamefont {Ghosh}\ \emph {et~al.}(2012)\citenamefont {Ghosh},
  \citenamefont {Mangum}, \citenamefont {Casson}, \citenamefont {Williams},
  \citenamefont {Htoon},\ and\ \citenamefont
  {Hollingsworth}}]{doi:10.1021/ja212032q}%
  \BibitemOpen
  \bibfield  {author} {\bibinfo {author} {\bibfnamefont {Yagnaseni}\
  \bibnamefont {Ghosh}}, \bibinfo {author} {\bibfnamefont {Benjamin~D.}\
  \bibnamefont {Mangum}}, \bibinfo {author} {\bibfnamefont {Joanna~L.}\
  \bibnamefont {Casson}}, \bibinfo {author} {\bibfnamefont {Darrick~J.}\
  \bibnamefont {Williams}}, \bibinfo {author} {\bibfnamefont {Han}\
  \bibnamefont {Htoon}}, \ and\ \bibinfo {author} {\bibfnamefont {Jennifer~A.}\
  \bibnamefont {Hollingsworth}},\ }\bibfield  {title} {\enquote {\bibinfo
  {title} {New insights into the complexities of shell growth and the strong
  influence of particle volume in nonblinking {``}giant{''} core/shell
  nanocrystal quantum dots},}\ }\href {\doibase 10.1021/ja212032q} {\bibfield
  {journal} {\bibinfo  {journal} {Journal of the American Chemical Society}\
  }\textbf {\bibinfo {volume} {134}},\ \bibinfo {pages} {9634--9643} (\bibinfo
  {year} {2012})}\BibitemShut {NoStop}%
\bibitem [{\citenamefont {Bae}\ \emph {et~al.}(2013)\citenamefont {Bae},
  \citenamefont {Padilha}, \citenamefont {Park}, \citenamefont {McDaniel},
  \citenamefont {Robel}, \citenamefont {Pietryga},\ and\ \citenamefont
  {Klimov}}]{doi:10.1021/nn4002825}%
  \BibitemOpen
  \bibfield  {author} {\bibinfo {author} {\bibfnamefont {Wan~Ki}\ \bibnamefont
  {Bae}}, \bibinfo {author} {\bibfnamefont {Lazaro~A.}\ \bibnamefont
  {Padilha}}, \bibinfo {author} {\bibfnamefont {Young-Shin}\ \bibnamefont
  {Park}}, \bibinfo {author} {\bibfnamefont {Hunter}\ \bibnamefont {McDaniel}},
  \bibinfo {author} {\bibfnamefont {Istvan}\ \bibnamefont {Robel}}, \bibinfo
  {author} {\bibfnamefont {Jeffrey~M.}\ \bibnamefont {Pietryga}}, \ and\
  \bibinfo {author} {\bibfnamefont {Victor~I.}\ \bibnamefont {Klimov}},\
  }\bibfield  {title} {\enquote {\bibinfo {title} {Controlled alloying of the
  core-shell interface in cdse/cds quantum dots for suppression of auger
  recombination},}\ }\href {\doibase 10.1021/nn4002825} {\bibfield  {journal}
  {\bibinfo  {journal} {ACS Nano}\ }\textbf {\bibinfo {volume} {7}},\ \bibinfo
  {pages} {3411--3419} (\bibinfo {year} {2013})}\BibitemShut {NoStop}%
\bibitem [{\citenamefont {Lin}\ \emph {et~al.}(2011)\citenamefont {Lin},
  \citenamefont {Franceschetti},\ and\ \citenamefont
  {Lusk}}]{doi:10.1021/nn200141f}%
  \BibitemOpen
  \bibfield  {author} {\bibinfo {author} {\bibfnamefont {Zhibin}\ \bibnamefont
  {Lin}}, \bibinfo {author} {\bibfnamefont {Alberto}\ \bibnamefont
  {Franceschetti}}, \ and\ \bibinfo {author} {\bibfnamefont {Mark~T.}\
  \bibnamefont {Lusk}},\ }\bibfield  {title} {\enquote {\bibinfo {title} {Size
  dependence of the multiple exciton generation rate in cdse quantum dots},}\
  }\href {\doibase 10.1021/nn200141f} {\bibfield  {journal} {\bibinfo
  {journal} {ACS Nano}\ }\textbf {\bibinfo {volume} {5}},\ \bibinfo {pages}
  {2503--2511} (\bibinfo {year} {2011})}\BibitemShut {NoStop}%
\bibitem [{\citenamefont {Alam}\ \emph {et~al.}(2012)\citenamefont {Alam},
  \citenamefont {Fontaine}, \citenamefont {Branchini},\ and\ \citenamefont
  {Maye}}]{doi:10.1021/nl301291g}%
  \BibitemOpen
  \bibfield  {author} {\bibinfo {author} {\bibfnamefont {Rabeka}\ \bibnamefont
  {Alam}}, \bibinfo {author} {\bibfnamefont {Danielle~M.}\ \bibnamefont
  {Fontaine}}, \bibinfo {author} {\bibfnamefont {Bruce~R.}\ \bibnamefont
  {Branchini}}, \ and\ \bibinfo {author} {\bibfnamefont {Mathew~M.}\
  \bibnamefont {Maye}},\ }\bibfield  {title} {\enquote {\bibinfo {title}
  {Designing quantum rods for optimized energy transfer with firefly luciferase
  enzymes},}\ }\href {\doibase 10.1021/nl301291g} {\bibfield  {journal}
  {\bibinfo  {journal} {Nano Letters}\ }\textbf {\bibinfo {volume} {12}},\
  \bibinfo {pages} {3251--3256} (\bibinfo {year} {2012})}\BibitemShut {NoStop}%
\bibitem [{\citenamefont {Han}\ \emph {et~al.}(2010)\citenamefont {Han},
  \citenamefont {Francesco},\ and\ \citenamefont
  {Maye}}]{doi:10.1021/jp107702b}%
  \BibitemOpen
  \bibfield  {author} {\bibinfo {author} {\bibfnamefont {Hyunjoo}\ \bibnamefont
  {Han}}, \bibinfo {author} {\bibfnamefont {Gianna~Di}\ \bibnamefont
  {Francesco}}, \ and\ \bibinfo {author} {\bibfnamefont {Mathew~M.}\
  \bibnamefont {Maye}},\ }\bibfield  {title} {\enquote {\bibinfo {title} {Size
  control and photophysical properties of quantum dots prepared via a novel
  tunable hydrothermal route},}\ }\href {\doibase 10.1021/jp107702b} {\bibfield
   {journal} {\bibinfo  {journal} {The Journal of Physical Chemistry C}\
  }\textbf {\bibinfo {volume} {114}},\ \bibinfo {pages} {19270--19277}
  (\bibinfo {year} {2010})}\BibitemShut {NoStop}%
\bibitem [{\citenamefont {Williams}\ \emph {et~al.}(2013)\citenamefont
  {Williams}, \citenamefont {Halliday}, \citenamefont {Mendis},\ and\
  \citenamefont {Durose}}]{0957-4484-24-13-135703}%
  \BibitemOpen
  \bibfield  {author} {\bibinfo {author} {\bibfnamefont {B~L}\ \bibnamefont
  {Williams}}, \bibinfo {author} {\bibfnamefont {D~P}\ \bibnamefont
  {Halliday}}, \bibinfo {author} {\bibfnamefont {B~G}\ \bibnamefont {Mendis}},
  \ and\ \bibinfo {author} {\bibfnamefont {K}~\bibnamefont {Durose}},\
  }\bibfield  {title} {\enquote {\bibinfo {title} {Microstructure and point
  defects in cdte nanowires for photovoltaic applications},}\ }\href {\doibase
  doi:10.1088/0957-4484/24/13/135703} {\bibfield  {journal} {\bibinfo
  {journal} {Nanotechnology}\ }\textbf {\bibinfo {volume} {24}},\ \bibinfo
  {pages} {135703} (\bibinfo {year} {2013})}\BibitemShut {NoStop}%
\bibitem [{\citenamefont {Yin}\ \emph {et~al.}(2013)\citenamefont {Yin},
  \citenamefont {Yue}, \citenamefont {Zang}, \citenamefont {Chiu},
  \citenamefont {Li}, \citenamefont {Kuo}, \citenamefont {Wu}, \citenamefont
  {Li}, \citenamefont {Fang},\ and\ \citenamefont {Chen}}]{C3NR00920C}%
  \BibitemOpen
  \bibfield  {author} {\bibinfo {author} {\bibfnamefont {Jun}\ \bibnamefont
  {Yin}}, \bibinfo {author} {\bibfnamefont {Chuang}\ \bibnamefont {Yue}},
  \bibinfo {author} {\bibfnamefont {Yashu}\ \bibnamefont {Zang}}, \bibinfo
  {author} {\bibfnamefont {Ching-Hsueh}\ \bibnamefont {Chiu}}, \bibinfo
  {author} {\bibfnamefont {Jinchai}\ \bibnamefont {Li}}, \bibinfo {author}
  {\bibfnamefont {Hao-Chung}\ \bibnamefont {Kuo}}, \bibinfo {author}
  {\bibfnamefont {Zhihao}\ \bibnamefont {Wu}}, \bibinfo {author} {\bibfnamefont
  {Jing}\ \bibnamefont {Li}}, \bibinfo {author} {\bibfnamefont {Yanyan}\
  \bibnamefont {Fang}}, \ and\ \bibinfo {author} {\bibfnamefont {Changqing}\
  \bibnamefont {Chen}},\ }\bibfield  {title} {\enquote {\bibinfo {title}
  {Effect of the surface-plasmon-exciton coupling and charge transfer process
  on the photoluminescence of metal-semiconductor nanostructures},}\ }\href
  {\doibase 10.1039/C3NR00920C} {\bibfield  {journal} {\bibinfo  {journal}
  {Nanoscale}\ }\textbf {\bibinfo {volume} {5}},\ \bibinfo {pages} {4436--4442}
  (\bibinfo {year} {2013})}\BibitemShut {NoStop}%
\bibitem [{\citenamefont {Jaeger}\ \emph {et~al.}(2012)\citenamefont {Jaeger},
  \citenamefont {Fischer},\ and\ \citenamefont {Prezhdo}}]{jaeger:064701}%
  \BibitemOpen
  \bibfield  {author} {\bibinfo {author} {\bibfnamefont {Heather~M.}\
  \bibnamefont {Jaeger}}, \bibinfo {author} {\bibfnamefont {Sean}\ \bibnamefont
  {Fischer}}, \ and\ \bibinfo {author} {\bibfnamefont {Oleg~V.}\ \bibnamefont
  {Prezhdo}},\ }\bibfield  {title} {\enquote {\bibinfo {title} {The role of
  surface defects in multi-exciton generation of lead selenide and silicon
  semiconductor quantum dots},}\ }\href {\doibase 10.1063/1.3682559} {\bibfield
   {journal} {\bibinfo  {journal} {The Journal of Chemical Physics}\ }\textbf
  {\bibinfo {volume} {136}},\ \bibinfo {eid} {064701} (\bibinfo {year}
  {2012})}\BibitemShut {NoStop}%
\bibitem [{\citenamefont {Kilina}\ \emph {et~al.}(2012)\citenamefont {Kilina},
  \citenamefont {Velizhanin}, \citenamefont {Ivanov}, \citenamefont {Prezhdo},\
  and\ \citenamefont {Tretiak}}]{doi:10.1021/nn302371q}%
  \BibitemOpen
  \bibfield  {author} {\bibinfo {author} {\bibfnamefont {Svetlana}\
  \bibnamefont {Kilina}}, \bibinfo {author} {\bibfnamefont {Kirill~A.}\
  \bibnamefont {Velizhanin}}, \bibinfo {author} {\bibfnamefont {Sergei}\
  \bibnamefont {Ivanov}}, \bibinfo {author} {\bibfnamefont {Oleg~V.}\
  \bibnamefont {Prezhdo}}, \ and\ \bibinfo {author} {\bibfnamefont {Sergei}\
  \bibnamefont {Tretiak}},\ }\bibfield  {title} {\enquote {\bibinfo {title}
  {Surface ligands increase photoexcitation relaxation rates in cdse quantum
  dots},}\ }\href {\doibase 10.1021/nn302371q} {\bibfield  {journal} {\bibinfo
  {journal} {ACS Nano}\ }\textbf {\bibinfo {volume} {6}},\ \bibinfo {pages}
  {6515--6524} (\bibinfo {year} {2012})}\BibitemShut {NoStop}%
\bibitem [{\citenamefont {Kilina}\ \emph
  {et~al.}(2009{\natexlab{a}})\citenamefont {Kilina}, \citenamefont {Ivanov},\
  and\ \citenamefont {Tretiak}}]{doi:10.1021/ja9005749}%
  \BibitemOpen
  \bibfield  {author} {\bibinfo {author} {\bibfnamefont {Svetlana}\
  \bibnamefont {Kilina}}, \bibinfo {author} {\bibfnamefont {Sergei}\
  \bibnamefont {Ivanov}}, \ and\ \bibinfo {author} {\bibfnamefont {Sergei}\
  \bibnamefont {Tretiak}},\ }\bibfield  {title} {\enquote {\bibinfo {title}
  {Effect of surface ligands on optical and electronic spectra of semiconductor
  nanoclusters},}\ }\href {\doibase 10.1021/ja9005749} {\bibfield  {journal}
  {\bibinfo  {journal} {Journal of the American Chemical Society}\ }\textbf
  {\bibinfo {volume} {131}},\ \bibinfo {pages} {7717--7726} (\bibinfo {year}
  {2009}{\natexlab{a}})}\BibitemShut {NoStop}%
\bibitem [{\citenamefont {Kilina}\ \emph
  {et~al.}(2009{\natexlab{b}})\citenamefont {Kilina}, \citenamefont {Kilin},\
  and\ \citenamefont {Prezhdo}}]{doi:10.1021/nn800674n}%
  \BibitemOpen
  \bibfield  {author} {\bibinfo {author} {\bibfnamefont {Svetlana~V.}\
  \bibnamefont {Kilina}}, \bibinfo {author} {\bibfnamefont {Dmitri~S.}\
  \bibnamefont {Kilin}}, \ and\ \bibinfo {author} {\bibfnamefont {Oleg~V.}\
  \bibnamefont {Prezhdo}},\ }\bibfield  {title} {\enquote {\bibinfo {title}
  {Breaking the phonon bottleneck in pbse and cdse quantum dots: Time-domain
  density functional theory of charge carrier relaxation},}\ }\href {\doibase
  10.1021/nn800674n} {\bibfield  {journal} {\bibinfo  {journal} {ACS Nano}\
  }\textbf {\bibinfo {volume} {3}},\ \bibinfo {pages} {93--99} (\bibinfo {year}
  {2009}{\natexlab{b}})}\BibitemShut {NoStop}%
\bibitem [{\citenamefont {Kelley}(2010)}]{Kelley20101296}%
  \BibitemOpen
  \bibfield  {author} {\bibinfo {author} {\bibfnamefont {A.M.}\ \bibnamefont
  {Kelley}},\ }\bibfield  {title} {\enquote {\bibinfo {title} {Electron-phonon
  coupling in cdse nanocrystals},}\ }\href {\doibase 10.1021/jz100123b}
  {\bibfield  {journal} {\bibinfo  {journal} {Journal of Physical Chemistry
  Letters}\ }\textbf {\bibinfo {volume} {1}},\ \bibinfo {pages} {1296--1300}
  (\bibinfo {year} {2010})}\BibitemShut {NoStop}%
\bibitem [{\citenamefont {Kelley}(2011)}]{Kelley20115254}%
  \BibitemOpen
  \bibfield  {author} {\bibinfo {author} {\bibfnamefont {A.M.}\ \bibnamefont
  {Kelley}},\ }\bibfield  {title} {\enquote {\bibinfo {title} {Electron-phonon
  coupling in cdse nanocrystals from an atomistic phonon model},}\ }\href
  {\doibase 10.1021/nn201475d} {\bibfield  {journal} {\bibinfo  {journal} {ACS
  Nano}\ }\textbf {\bibinfo {volume} {5}},\ \bibinfo {pages} {5254--5262}
  (\bibinfo {year} {2011})}\BibitemShut {NoStop}%
\bibitem [{\citenamefont {Kelley}\ \emph {et~al.}(2012)\citenamefont {Kelley},
  \citenamefont {Dai}, \citenamefont {Jiang}, \citenamefont {Baker},\ and\
  \citenamefont {Kelley}}]{Kelley2012}%
  \BibitemOpen
  \bibfield  {author} {\bibinfo {author} {\bibfnamefont {A.M.}\ \bibnamefont
  {Kelley}}, \bibinfo {author} {\bibfnamefont {Q.}~\bibnamefont {Dai}},
  \bibinfo {author} {\bibfnamefont {Z.-j.}\ \bibnamefont {Jiang}}, \bibinfo
  {author} {\bibfnamefont {J.A.}\ \bibnamefont {Baker}}, \ and\ \bibinfo
  {author} {\bibfnamefont {D.F.}\ \bibnamefont {Kelley}},\ }\bibfield  {title}
  {\enquote {\bibinfo {title} {Resonance raman spectra of wurtzite and
  zincblende cdse nanocrystals},}\ }\href {\doibase
  10.1016/j.chemphys.2012.09.029} {\bibfield  {journal} {\bibinfo  {journal}
  {Chemical Physics}\ } (\bibinfo {year} {2012}),\
  10.1016/j.chemphys.2012.09.029}\BibitemShut {NoStop}%
\bibitem [{\citenamefont {Kilina}\ \emph {et~al.}(2011)\citenamefont {Kilina},
  \citenamefont {Kilin}, \citenamefont {Prezhdo},\ and\ \citenamefont
  {Prezhdo}}]{Kilina201121641}%
  \BibitemOpen
  \bibfield  {author} {\bibinfo {author} {\bibfnamefont {S.V.}\ \bibnamefont
  {Kilina}}, \bibinfo {author} {\bibfnamefont {D.S.}\ \bibnamefont {Kilin}},
  \bibinfo {author} {\bibfnamefont {V.V.}\ \bibnamefont {Prezhdo}}, \ and\
  \bibinfo {author} {\bibfnamefont {O.V.}\ \bibnamefont {Prezhdo}},\ }\bibfield
   {title} {\enquote {\bibinfo {title} {Theoretical study of electron-phonon
  relaxation in pbse and cdse quantum dots: Evidence for phonon memory},}\
  }\href {\doibase 10.1021/jp206594e} {\bibfield  {journal} {\bibinfo
  {journal} {Journal of Physical Chemistry C}\ }\textbf {\bibinfo {volume}
  {115}},\ \bibinfo {pages} {21641--21651} (\bibinfo {year}
  {2011})}\BibitemShut {NoStop}%
\bibitem [{\citenamefont {Hyeon-Deuk}\ and\ \citenamefont
  {Prezhdo}(2011)}]{Hyeon-Deuk20111845}%
  \BibitemOpen
  \bibfield  {author} {\bibinfo {author} {\bibfnamefont {K.}~\bibnamefont
  {Hyeon-Deuk}}\ and\ \bibinfo {author} {\bibfnamefont {O.V.}\ \bibnamefont
  {Prezhdo}},\ }\bibfield  {title} {\enquote {\bibinfo {title} {Time-domain ab
  initio study of auger and phonon-assisted auger processes in a semiconductor
  quantum dot},}\ }\href {\doibase 10.1021/nl200651p} {\bibfield  {journal}
  {\bibinfo  {journal} {Nano Letters}\ }\textbf {\bibinfo {volume} {11}},\
  \bibinfo {pages} {1845--1850} (\bibinfo {year} {2011})}\BibitemShut {NoStop}%
\bibitem [{\citenamefont {Hyeon-Deuk}\ and\ \citenamefont
  {Prezhdo}(2012{\natexlab{a}})}]{Hyeon-Deuk2012}%
  \BibitemOpen
  \bibfield  {author} {\bibinfo {author} {\bibfnamefont {K.}~\bibnamefont
  {Hyeon-Deuk}}\ and\ \bibinfo {author} {\bibfnamefont {O.V.}\ \bibnamefont
  {Prezhdo}},\ }\bibfield  {title} {\enquote {\bibinfo {title} {Photoexcited
  electron and hole dynamics in semiconductor quantum dots: Phonon-induced
  relaxation, dephasing, multiple exciton generation and recombination},}\
  }\href {\doibase 10.1088/0953-8984/24/36/363201} {\bibfield  {journal}
  {\bibinfo  {journal} {Journal of Physics Condensed Matter}\ }\textbf
  {\bibinfo {volume} {24}} (\bibinfo {year} {2012}{\natexlab{a}}),\
  10.1088/0953-8984/24/36/363201}\BibitemShut {NoStop}%
\bibitem [{\citenamefont {Kilina}\ \emph {et~al.}(2013)\citenamefont {Kilina},
  \citenamefont {Neukirch}, \citenamefont {Habenicht}, \citenamefont {Kilin},\
  and\ \citenamefont {Prezhdo}}]{Kilina2013}%
  \BibitemOpen
  \bibfield  {author} {\bibinfo {author} {\bibfnamefont {S.V.}\ \bibnamefont
  {Kilina}}, \bibinfo {author} {\bibfnamefont {A.J.}\ \bibnamefont {Neukirch}},
  \bibinfo {author} {\bibfnamefont {B.F.}\ \bibnamefont {Habenicht}}, \bibinfo
  {author} {\bibfnamefont {D.S.}\ \bibnamefont {Kilin}}, \ and\ \bibinfo
  {author} {\bibfnamefont {O.V.}\ \bibnamefont {Prezhdo}},\ }\bibfield  {title}
  {\enquote {\bibinfo {title} {Quantum zeno effect rationalizes the phonon
  bottleneck in semiconductor quantum dots},}\ }\href {\doibase
  10.1103/PhysRevLett.110.180404} {\bibfield  {journal} {\bibinfo  {journal}
  {Physical Review Letters}\ }\textbf {\bibinfo {volume} {110}} (\bibinfo
  {year} {2013}),\ 10.1103/PhysRevLett.110.180404}\BibitemShut {NoStop}%
\bibitem [{\citenamefont {Musa}\ \emph {et~al.}(2011)\citenamefont {Musa},
  \citenamefont {Massuyeau}, \citenamefont {Cario}, \citenamefont {Duvail},
  \citenamefont {Jobic}, \citenamefont {Deniard},\ and\ \citenamefont
  {Faulques}}]{musa:243107}%
  \BibitemOpen
  \bibfield  {author} {\bibinfo {author} {\bibfnamefont {I.}~\bibnamefont
  {Musa}}, \bibinfo {author} {\bibfnamefont {F.}~\bibnamefont {Massuyeau}},
  \bibinfo {author} {\bibfnamefont {L.}~\bibnamefont {Cario}}, \bibinfo
  {author} {\bibfnamefont {J.~L.}\ \bibnamefont {Duvail}}, \bibinfo {author}
  {\bibfnamefont {S.}~\bibnamefont {Jobic}}, \bibinfo {author} {\bibfnamefont
  {P.}~\bibnamefont {Deniard}}, \ and\ \bibinfo {author} {\bibfnamefont
  {E.}~\bibnamefont {Faulques}},\ }\bibfield  {title} {\enquote {\bibinfo
  {title} {Temperature and size dependence of time-resolved exciton
  recombination in zno quantum dots},}\ }\href {\doibase 10.1063/1.3669511}
  {\bibfield  {journal} {\bibinfo  {journal} {Applied Physics Letters}\
  }\textbf {\bibinfo {volume} {99}},\ \bibinfo {eid} {243107} (\bibinfo {year}
  {2011})}\BibitemShut {NoStop}%
\bibitem [{\citenamefont {Chon}\ \emph {et~al.}(2011)\citenamefont {Chon},
  \citenamefont {Bang}, \citenamefont {Park}, \citenamefont {Jeong},
  \citenamefont {Choi}, \citenamefont {Lee}, \citenamefont {Joo},\ and\
  \citenamefont {Kim}}]{doi:10.1021/jp109229u}%
  \BibitemOpen
  \bibfield  {author} {\bibinfo {author} {\bibfnamefont {Bonghwan}\
  \bibnamefont {Chon}}, \bibinfo {author} {\bibfnamefont {Jiwon}\ \bibnamefont
  {Bang}}, \bibinfo {author} {\bibfnamefont {Juwon}\ \bibnamefont {Park}},
  \bibinfo {author} {\bibfnamefont {Cherlhyun}\ \bibnamefont {Jeong}}, \bibinfo
  {author} {\bibfnamefont {Jong~Hwa}\ \bibnamefont {Choi}}, \bibinfo {author}
  {\bibfnamefont {Jong-Bong}\ \bibnamefont {Lee}}, \bibinfo {author}
  {\bibfnamefont {Taiha}\ \bibnamefont {Joo}}, \ and\ \bibinfo {author}
  {\bibfnamefont {Sungjee}\ \bibnamefont {Kim}},\ }\bibfield  {title} {\enquote
  {\bibinfo {title} {Unique temperature dependence and blinking behavior of
  cdte/cdse (core/shell) type-ii quantum dots},}\ }\href {\doibase
  10.1021/jp109229u} {\bibfield  {journal} {\bibinfo  {journal} {The Journal of
  Physical Chemistry C}\ }\textbf {\bibinfo {volume} {115}},\ \bibinfo {pages}
  {436--442} (\bibinfo {year} {2011})}\BibitemShut {NoStop}%
\bibitem [{\citenamefont {Franceschetti}\ and\ \citenamefont
  {Zunger}(2001)}]{PhysRevB.63.153304}%
  \BibitemOpen
  \bibfield  {author} {\bibinfo {author} {\bibfnamefont {Alberto}\ \bibnamefont
  {Franceschetti}}\ and\ \bibinfo {author} {\bibfnamefont {Alex}\ \bibnamefont
  {Zunger}},\ }\bibfield  {title} {\enquote {\bibinfo {title} {Exciton
  dissociation and interdot transport in cdse quantum-dot molecules},}\ }\href
  {\doibase 10.1103/PhysRevB.63.153304} {\bibfield  {journal} {\bibinfo
  {journal} {Phys. Rev. B}\ }\textbf {\bibinfo {volume} {63}},\ \bibinfo
  {pages} {153304} (\bibinfo {year} {2001})}\BibitemShut {NoStop}%
\bibitem [{\citenamefont {Zhu}\ \emph {et~al.}(2011)\citenamefont {Zhu},
  \citenamefont {Song},\ and\ \citenamefont {Lian}}]{doi:10.1021/ja202752s}%
  \BibitemOpen
  \bibfield  {author} {\bibinfo {author} {\bibfnamefont {Haiming}\ \bibnamefont
  {Zhu}}, \bibinfo {author} {\bibfnamefont {Nianhui}\ \bibnamefont {Song}}, \
  and\ \bibinfo {author} {\bibfnamefont {Tianquan}\ \bibnamefont {Lian}},\
  }\bibfield  {title} {\enquote {\bibinfo {title} {Wave function engineering
  for ultrafast charge separation and slow charge recombination in type ii
  core/shell quantum dots},}\ }\href {\doibase 10.1021/ja202752s} {\bibfield
  {journal} {\bibinfo  {journal} {Journal of the American Chemical Society}\
  }\textbf {\bibinfo {volume} {133}},\ \bibinfo {pages} {8762--8771} (\bibinfo
  {year} {2011})}\BibitemShut {NoStop}%
\bibitem [{\citenamefont {Zhu}\ \emph {et~al.}(2010)\citenamefont {Zhu},
  \citenamefont {Song},\ and\ \citenamefont {Lian}}]{doi:10.1021/ja106710m}%
  \BibitemOpen
  \bibfield  {author} {\bibinfo {author} {\bibfnamefont {Haiming}\ \bibnamefont
  {Zhu}}, \bibinfo {author} {\bibfnamefont {Nianhui}\ \bibnamefont {Song}}, \
  and\ \bibinfo {author} {\bibfnamefont {Tianquan}\ \bibnamefont {Lian}},\
  }\bibfield  {title} {\enquote {\bibinfo {title} {Controlling charge
  separation and recombination rates in cdse/zns type i core-shell quantum dots
  by shell thicknesses},}\ }\href {\doibase 10.1021/ja106710m} {\bibfield
  {journal} {\bibinfo  {journal} {Journal of the American Chemical Society}\
  }\textbf {\bibinfo {volume} {132}},\ \bibinfo {pages} {15038--15045}
  (\bibinfo {year} {2010})}\BibitemShut {NoStop}%
\bibitem [{\citenamefont {Hoy}\ \emph {et~al.}(0)\citenamefont {Hoy},
  \citenamefont {Morrison}, \citenamefont {Steinberg}, \citenamefont {Buhro},\
  and\ \citenamefont {Loomis}}]{doi:10.1021/jz4004735}%
  \BibitemOpen
  \bibfield  {author} {\bibinfo {author} {\bibfnamefont {Jessica}\ \bibnamefont
  {Hoy}}, \bibinfo {author} {\bibfnamefont {Paul~J.}\ \bibnamefont {Morrison}},
  \bibinfo {author} {\bibfnamefont {Lindsey~K.}\ \bibnamefont {Steinberg}},
  \bibinfo {author} {\bibfnamefont {William~E.}\ \bibnamefont {Buhro}}, \ and\
  \bibinfo {author} {\bibfnamefont {Richard~A.}\ \bibnamefont {Loomis}},\
  }\bibfield  {title} {\enquote {\bibinfo {title} {Excitation energy dependence
  of the photoluminescence quantum yields of core and core/shell quantum
  dots},}\ }\href {\doibase 10.1021/jz4004735} {\bibfield  {journal} {\bibinfo
  {journal} {The Journal of Physical Chemistry Letters}\ }\textbf {\bibinfo
  {volume} {0}},\ \bibinfo {pages} {2053--2060} (\bibinfo {year}
  {0})}\BibitemShut {NoStop}%
\bibitem [{\citenamefont {Li}\ \emph {et~al.}(2012)\citenamefont {Li},
  \citenamefont {Zhang}, \citenamefont {Zhang},\ and\ \citenamefont
  {Xiong}}]{doi:10.1021/nl300749z}%
  \BibitemOpen
  \bibfield  {author} {\bibinfo {author} {\bibfnamefont {Dehui}\ \bibnamefont
  {Li}}, \bibinfo {author} {\bibfnamefont {Jun}\ \bibnamefont {Zhang}},
  \bibinfo {author} {\bibfnamefont {Qing}\ \bibnamefont {Zhang}}, \ and\
  \bibinfo {author} {\bibfnamefont {Qihua}\ \bibnamefont {Xiong}},\ }\bibfield
  {title} {\enquote {\bibinfo {title} {Electric-field-dependent
  photoconductivity in cds nanowires and nanobelts: Exciton ionization,
  franz-keldysh, and stark effects},}\ }\href {\doibase 10.1021/nl300749z}
  {\bibfield  {journal} {\bibinfo  {journal} {Nano Letters}\ }\textbf {\bibinfo
  {volume} {12}},\ \bibinfo {pages} {2993--2999} (\bibinfo {year}
  {2012})}\BibitemShut {NoStop}%
\bibitem [{\citenamefont {Yaacobi-Gross}\ \emph {et~al.}(2011)\citenamefont
  {Yaacobi-Gross}, \citenamefont {Soreni-Harari}, \citenamefont {Zimin},
  \citenamefont {Kababya}, \citenamefont {Schmidt},\ and\ \citenamefont
  {Tessler}}]{yaacobi2011molecular}%
  \BibitemOpen
  \bibfield  {author} {\bibinfo {author} {\bibfnamefont {Nir}\ \bibnamefont
  {Yaacobi-Gross}}, \bibinfo {author} {\bibfnamefont {Michal}\ \bibnamefont
  {Soreni-Harari}}, \bibinfo {author} {\bibfnamefont {Marina}\ \bibnamefont
  {Zimin}}, \bibinfo {author} {\bibfnamefont {Shifi}\ \bibnamefont {Kababya}},
  \bibinfo {author} {\bibfnamefont {Asher}\ \bibnamefont {Schmidt}}, \ and\
  \bibinfo {author} {\bibfnamefont {Nir}\ \bibnamefont {Tessler}},\ }\bibfield
  {title} {\enquote {\bibinfo {title} {Molecular control of quantum-dot
  internal electric field and its application to cdse-based solar cells},}\
  }\href {\doibase 10.1038/nmat3133} {\bibfield  {journal} {\bibinfo  {journal}
  {Nature Materials}\ }\textbf {\bibinfo {volume} {10}},\ \bibinfo {pages}
  {974--979} (\bibinfo {year} {2011})}\BibitemShut {NoStop}%
\bibitem [{\citenamefont {Liu}\ \emph {et~al.}(2012)\citenamefont {Liu},
  \citenamefont {Borys}, \citenamefont {Huang}, \citenamefont {Talapin},\ and\
  \citenamefont {Lupton}}]{PhysRevB.86.045303}%
  \BibitemOpen
  \bibfield  {author} {\bibinfo {author} {\bibfnamefont {Su}~\bibnamefont
  {Liu}}, \bibinfo {author} {\bibfnamefont {Nicholas~J.}\ \bibnamefont
  {Borys}}, \bibinfo {author} {\bibfnamefont {Jing}\ \bibnamefont {Huang}},
  \bibinfo {author} {\bibfnamefont {Dmitri~V.}\ \bibnamefont {Talapin}}, \ and\
  \bibinfo {author} {\bibfnamefont {John~M.}\ \bibnamefont {Lupton}},\
  }\bibfield  {title} {\enquote {\bibinfo {title} {Exciton storage in cdse/cds
  tetrapod semiconductor nanocrystals: Electric field effects on exciton and
  multiexciton states},}\ }\href {\doibase 10.1103/PhysRevB.86.045303}
  {\bibfield  {journal} {\bibinfo  {journal} {Phys. Rev. B}\ }\textbf {\bibinfo
  {volume} {86}},\ \bibinfo {pages} {045303} (\bibinfo {year}
  {2012})}\BibitemShut {NoStop}%
\bibitem [{\citenamefont {Perebeinos}\ and\ \citenamefont
  {Avouris}(2007)}]{doi:10.1021/nl0625022}%
  \BibitemOpen
  \bibfield  {author} {\bibinfo {author} {\bibfnamefont {Vasili}\ \bibnamefont
  {Perebeinos}}\ and\ \bibinfo {author} {\bibfnamefont {Phaedon}\ \bibnamefont
  {Avouris}},\ }\bibfield  {title} {\enquote {\bibinfo {title} {Exciton
  ionization, franz-keldysh, and stark effects in carbon nanotubes},}\ }\href
  {\doibase 10.1021/nl0625022} {\bibfield  {journal} {\bibinfo  {journal} {Nano
  Letters}\ }\textbf {\bibinfo {volume} {7}},\ \bibinfo {pages} {609--613}
  (\bibinfo {year} {2007})}\BibitemShut {NoStop}%
\bibitem [{\citenamefont {Blanton}\ \emph {et~al.}(2013)\citenamefont
  {Blanton}, \citenamefont {Brenon},\ and\ \citenamefont
  {Chakraborty}}]{blanton:054114}%
  \BibitemOpen
  \bibfield  {author} {\bibinfo {author} {\bibfnamefont {Christopher~J.}\
  \bibnamefont {Blanton}}, \bibinfo {author} {\bibfnamefont {Christopher}\
  \bibnamefont {Brenon}}, \ and\ \bibinfo {author} {\bibfnamefont {Arindam}\
  \bibnamefont {Chakraborty}},\ }\bibfield  {title} {\enquote {\bibinfo {title}
  {Development of polaron-transformed explicitly correlated full configuration
  interaction method for investigation of quantum-confined stark effect in gaas
  quantum dots},}\ }\href {\doibase 10.1063/1.4789540} {\bibfield  {journal}
  {\bibinfo  {journal} {The Journal of Chemical Physics}\ }\textbf {\bibinfo
  {volume} {138}},\ \bibinfo {eid} {054114} (\bibinfo {year}
  {2013})}\BibitemShut {NoStop}%
\bibitem [{\citenamefont {Park}\ \emph {et~al.}(2012)\citenamefont {Park},
  \citenamefont {Deutsch}, \citenamefont {Li}, \citenamefont {Oron},\ and\
  \citenamefont {Weiss}}]{doi:10.1021/nn303719m}%
  \BibitemOpen
  \bibfield  {author} {\bibinfo {author} {\bibfnamefont {KyoungWon}\
  \bibnamefont {Park}}, \bibinfo {author} {\bibfnamefont {Zvicka}\ \bibnamefont
  {Deutsch}}, \bibinfo {author} {\bibfnamefont {J.~Jack}\ \bibnamefont {Li}},
  \bibinfo {author} {\bibfnamefont {Dan}\ \bibnamefont {Oron}}, \ and\ \bibinfo
  {author} {\bibfnamefont {Shimon}\ \bibnamefont {Weiss}},\ }\bibfield  {title}
  {\enquote {\bibinfo {title} {Single molecule quantum-confined stark effect
  measurements of semiconductor nanoparticles at room temperature},}\ }\href
  {\doibase 10.1021/nn303719m} {\bibfield  {journal} {\bibinfo  {journal} {ACS
  Nano}\ }\textbf {\bibinfo {volume} {6}},\ \bibinfo {pages} {10013--10023}
  (\bibinfo {year} {2012})}\BibitemShut {NoStop}%
\bibitem [{\citenamefont {Franceschetti}\ and\ \citenamefont
  {Zunger}(1997)}]{PhysRevLett.78.915}%
  \BibitemOpen
  \bibfield  {author} {\bibinfo {author} {\bibfnamefont {Alberto}\ \bibnamefont
  {Franceschetti}}\ and\ \bibinfo {author} {\bibfnamefont {Alex}\ \bibnamefont
  {Zunger}},\ }\bibfield  {title} {\enquote {\bibinfo {title} {Direct
  pseudopotential calculation of exciton coulomb and exchange energies in
  semiconductor quantum dots},}\ }\href {\doibase 10.1103/PhysRevLett.78.915}
  {\bibfield  {journal} {\bibinfo  {journal} {Phys. Rev. Lett.}\ }\textbf
  {\bibinfo {volume} {78}},\ \bibinfo {pages} {915--918} (\bibinfo {year}
  {1997})}\BibitemShut {NoStop}%
\bibitem [{\citenamefont {Meulenberg}\ \emph {et~al.}(2009)\citenamefont
  {Meulenberg}, \citenamefont {Lee}, \citenamefont {Wolcott}, \citenamefont
  {Zhang}, \citenamefont {Terminello},\ and\ \citenamefont {van
  Buuren}}]{doi:10.1021/nn8006916}%
  \BibitemOpen
  \bibfield  {author} {\bibinfo {author} {\bibfnamefont {Robert~W.}\
  \bibnamefont {Meulenberg}}, \bibinfo {author} {\bibfnamefont {Jonathan~R.I.}\
  \bibnamefont {Lee}}, \bibinfo {author} {\bibfnamefont {Abraham}\ \bibnamefont
  {Wolcott}}, \bibinfo {author} {\bibfnamefont {Jin~Z.}\ \bibnamefont {Zhang}},
  \bibinfo {author} {\bibfnamefont {Louis~J.}\ \bibnamefont {Terminello}}, \
  and\ \bibinfo {author} {\bibfnamefont {Tony}\ \bibnamefont {van Buuren}},\
  }\bibfield  {title} {\enquote {\bibinfo {title} {Determination of the exciton
  binding energy in cdse quantum dots},}\ }\href {\doibase 10.1021/nn8006916}
  {\bibfield  {journal} {\bibinfo  {journal} {ACS Nano}\ }\textbf {\bibinfo
  {volume} {3}},\ \bibinfo {pages} {325--330} (\bibinfo {year}
  {2009})}\BibitemShut {NoStop}%
\bibitem [{\citenamefont {Jasieniak}\ \emph {et~al.}(2011)\citenamefont
  {Jasieniak}, \citenamefont {Califano},\ and\ \citenamefont
  {Watkins}}]{doi:10.1021/nn201681s}%
  \BibitemOpen
  \bibfield  {author} {\bibinfo {author} {\bibfnamefont {Jacek}\ \bibnamefont
  {Jasieniak}}, \bibinfo {author} {\bibfnamefont {Marco}\ \bibnamefont
  {Califano}}, \ and\ \bibinfo {author} {\bibfnamefont {Scott~E.}\ \bibnamefont
  {Watkins}},\ }\bibfield  {title} {\enquote {\bibinfo {title} {Size-dependent
  valence and conduction band-edge energies of semiconductor nanocrystals},}\
  }\href {\doibase 10.1021/nn201681s} {\bibfield  {journal} {\bibinfo
  {journal} {ACS Nano}\ }\textbf {\bibinfo {volume} {5}},\ \bibinfo {pages}
  {5888--5902} (\bibinfo {year} {2011})}\BibitemShut {NoStop}%
\bibitem [{\citenamefont {Maan}\ \emph {et~al.}(1984)\citenamefont {Maan},
  \citenamefont {Belle}, \citenamefont {Fasolino}, \citenamefont {Altarelli},\
  and\ \citenamefont {Ploog}}]{PhysRevB.30.2253}%
  \BibitemOpen
  \bibfield  {author} {\bibinfo {author} {\bibfnamefont {J.~C.}\ \bibnamefont
  {Maan}}, \bibinfo {author} {\bibfnamefont {G.}~\bibnamefont {Belle}},
  \bibinfo {author} {\bibfnamefont {A.}~\bibnamefont {Fasolino}}, \bibinfo
  {author} {\bibfnamefont {M.}~\bibnamefont {Altarelli}}, \ and\ \bibinfo
  {author} {\bibfnamefont {K.}~\bibnamefont {Ploog}},\ }\bibfield  {title}
  {\enquote {\bibinfo {title} {Magneto-optical determination of exciton binding
  energy in gaas-${\mathrm{ga}}_{1-x}{\mathrm{al}}_{x}\mathrm{As}$ quantum
  wells},}\ }\href {\doibase 10.1103/PhysRevB.30.2253} {\bibfield  {journal}
  {\bibinfo  {journal} {Phys. Rev. B}\ }\textbf {\bibinfo {volume} {30}},\
  \bibinfo {pages} {2253--2256} (\bibinfo {year} {1984})}\BibitemShut {NoStop}%
\bibitem [{\citenamefont {Ramvall}\ \emph {et~al.}(1998)\citenamefont
  {Ramvall}, \citenamefont {Tanaka}, \citenamefont {Nomura}, \citenamefont
  {Riblet},\ and\ \citenamefont {Aoyagi}}]{ramvall:1104}%
  \BibitemOpen
  \bibfield  {author} {\bibinfo {author} {\bibfnamefont {Peter}\ \bibnamefont
  {Ramvall}}, \bibinfo {author} {\bibfnamefont {Satoru}\ \bibnamefont
  {Tanaka}}, \bibinfo {author} {\bibfnamefont {Shintaro}\ \bibnamefont
  {Nomura}}, \bibinfo {author} {\bibfnamefont {Philippe}\ \bibnamefont
  {Riblet}}, \ and\ \bibinfo {author} {\bibfnamefont {Yoshinobu}\ \bibnamefont
  {Aoyagi}},\ }\bibfield  {title} {\enquote {\bibinfo {title} {Observation of
  confinement-dependent exciton binding energy of gan quantum dots},}\ }\href
  {\doibase 10.1063/1.122098} {\bibfield  {journal} {\bibinfo  {journal}
  {Applied Physics Letters}\ }\textbf {\bibinfo {volume} {73}},\ \bibinfo
  {pages} {1104--1106} (\bibinfo {year} {1998})}\BibitemShut {NoStop}%
\bibitem [{\citenamefont {Wang}\ and\ \citenamefont
  {Zunger}(1996)}]{PhysRevB.53.9579}%
  \BibitemOpen
  \bibfield  {author} {\bibinfo {author} {\bibfnamefont {Lin-Wang}\
  \bibnamefont {Wang}}\ and\ \bibinfo {author} {\bibfnamefont {Alex}\
  \bibnamefont {Zunger}},\ }\bibfield  {title} {\enquote {\bibinfo {title}
  {Pseudopotential calculations of nanoscale cdse quantum dots},}\ }\href
  {\doibase 10.1103/PhysRevB.53.9579} {\bibfield  {journal} {\bibinfo
  {journal} {Phys. Rev. B}\ }\textbf {\bibinfo {volume} {53}},\ \bibinfo
  {pages} {9579--9582} (\bibinfo {year} {1996})}\BibitemShut {NoStop}%
\bibitem [{\citenamefont {Kucur}\ \emph {et~al.}(2003)\citenamefont {Kucur},
  \citenamefont {Riegler}, \citenamefont {Urban},\ and\ \citenamefont
  {Nann}}]{kucur:2333}%
  \BibitemOpen
  \bibfield  {author} {\bibinfo {author} {\bibfnamefont {Erol}\ \bibnamefont
  {Kucur}}, \bibinfo {author} {\bibfnamefont {Jurgen}\ \bibnamefont {Riegler}},
  \bibinfo {author} {\bibfnamefont {Gerald~A.}\ \bibnamefont {Urban}}, \ and\
  \bibinfo {author} {\bibfnamefont {Thomas}\ \bibnamefont {Nann}},\ }\bibfield
  {title} {\enquote {\bibinfo {title} {Determination of quantum confinement in
  cdse nanocrystals by cyclic voltammetry},}\ }\href {\doibase
  10.1063/1.1582834} {\bibfield  {journal} {\bibinfo  {journal} {The Journal of
  Chemical Physics}\ }\textbf {\bibinfo {volume} {119}},\ \bibinfo {pages}
  {2333--2337} (\bibinfo {year} {2003})}\BibitemShut {NoStop}%
\bibitem [{\citenamefont {Inamdar}\ \emph {et~al.}(2008)\citenamefont
  {Inamdar}, \citenamefont {Ingole},\ and\ \citenamefont
  {Haram}}]{CPHC:CPHC200800482}%
  \BibitemOpen
  \bibfield  {author} {\bibinfo {author} {\bibfnamefont {Shaukatali~N.}\
  \bibnamefont {Inamdar}}, \bibinfo {author} {\bibfnamefont {Pravin~P.}\
  \bibnamefont {Ingole}}, \ and\ \bibinfo {author} {\bibfnamefont {Santosh~K.}\
  \bibnamefont {Haram}},\ }\bibfield  {title} {\enquote {\bibinfo {title}
  {Determination of band structure parameters and the quasi-particle gap of
  cdse quantum dots by cyclic voltammetry},}\ }\href {\doibase
  10.1002/cphc.200800482} {\bibfield  {journal} {\bibinfo  {journal}
  {ChemPhysChem}\ }\textbf {\bibinfo {volume} {9}},\ \bibinfo {pages}
  {2574--2579} (\bibinfo {year} {2008})}\BibitemShut {NoStop}%
\bibitem [{\citenamefont {Querner}\ \emph {et~al.}(2005)\citenamefont
  {Querner}, \citenamefont {Reiss}, \citenamefont {Sadki}, \citenamefont
  {Zagorska},\ and\ \citenamefont {Pron}}]{B508268B}%
  \BibitemOpen
  \bibfield  {author} {\bibinfo {author} {\bibfnamefont {Claudia}\ \bibnamefont
  {Querner}}, \bibinfo {author} {\bibfnamefont {Peter}\ \bibnamefont {Reiss}},
  \bibinfo {author} {\bibfnamefont {Said}\ \bibnamefont {Sadki}}, \bibinfo
  {author} {\bibfnamefont {Malgorzata}\ \bibnamefont {Zagorska}}, \ and\
  \bibinfo {author} {\bibfnamefont {Adam}\ \bibnamefont {Pron}},\ }\bibfield
  {title} {\enquote {\bibinfo {title} {Size and ligand effects on the
  electrochemical and spectroelectrochemical responses of cdse nanocrystals},}\
  }\href {\doibase 10.1039/B508268B} {\bibfield  {journal} {\bibinfo  {journal}
  {Phys. Chem. Chem. Phys.}\ }\textbf {\bibinfo {volume} {7}},\ \bibinfo
  {pages} {3204--3209} (\bibinfo {year} {2005})}\BibitemShut {NoStop}%
\bibitem [{\citenamefont {Garc{\'i}a-Santamar{\'i}a}\ \emph
  {et~al.}(2009)\citenamefont {Garc{\'i}a-Santamar{\'i}a}, \citenamefont
  {Chen}, \citenamefont {Vela}, \citenamefont {Schaller}, \citenamefont
  {Hollingsworth},\ and\ \citenamefont {Klimov}}]{doi:10.1021/nl901681d}%
  \BibitemOpen
  \bibfield  {author} {\bibinfo {author} {\bibfnamefont {Florencio}\
  \bibnamefont {Garc{\'i}a-Santamar{\'i}a}}, \bibinfo {author} {\bibfnamefont
  {Yongfen}\ \bibnamefont {Chen}}, \bibinfo {author} {\bibfnamefont {Javier}\
  \bibnamefont {Vela}}, \bibinfo {author} {\bibfnamefont {Richard~D.}\
  \bibnamefont {Schaller}}, \bibinfo {author} {\bibfnamefont {Jennifer~A.}\
  \bibnamefont {Hollingsworth}}, \ and\ \bibinfo {author} {\bibfnamefont
  {Victor~I.}\ \bibnamefont {Klimov}},\ }\bibfield  {title} {\enquote {\bibinfo
  {title} {Suppressed auger recombination in {``}giant{''} nanocrystals boosts
  optical gain performance},}\ }\href {\doibase 10.1021/nl901681d} {\bibfield
  {journal} {\bibinfo  {journal} {Nano Letters}\ }\textbf {\bibinfo {volume}
  {9}},\ \bibinfo {pages} {3482--3488} (\bibinfo {year} {2009})}\BibitemShut
  {NoStop}%
\bibitem [{\citenamefont {Wang}\ \emph {et~al.}(2003)\citenamefont {Wang},
  \citenamefont {Califano}, \citenamefont {Zunger},\ and\ \citenamefont
  {Franceschetti}}]{PhysRevLett.91.056404}%
  \BibitemOpen
  \bibfield  {author} {\bibinfo {author} {\bibfnamefont {Lin-Wang}\
  \bibnamefont {Wang}}, \bibinfo {author} {\bibfnamefont {Marco}\ \bibnamefont
  {Califano}}, \bibinfo {author} {\bibfnamefont {Alex}\ \bibnamefont {Zunger}},
  \ and\ \bibinfo {author} {\bibfnamefont {Alberto}\ \bibnamefont
  {Franceschetti}},\ }\bibfield  {title} {\enquote {\bibinfo {title}
  {Pseudopotential theory of auger processes in cdse quantum dots},}\ }\href
  {\doibase 10.1103/PhysRevLett.91.056404} {\bibfield  {journal} {\bibinfo
  {journal} {Phys. Rev. Lett.}\ }\textbf {\bibinfo {volume} {91}},\ \bibinfo
  {pages} {056404} (\bibinfo {year} {2003})}\BibitemShut {NoStop}%
\bibitem [{\citenamefont {Hyeon-Deuk}\ and\ \citenamefont
  {Prezhdo}(2012{\natexlab{b}})}]{doi:10.1021/nn2038884}%
  \BibitemOpen
  \bibfield  {author} {\bibinfo {author} {\bibfnamefont {Kim}\ \bibnamefont
  {Hyeon-Deuk}}\ and\ \bibinfo {author} {\bibfnamefont {Oleg~V.}\ \bibnamefont
  {Prezhdo}},\ }\bibfield  {title} {\enquote {\bibinfo {title} {Multiple
  exciton generation and recombination dynamics in small si and cdse quantum
  dots: An ab initio time-domain study},}\ }\href {\doibase 10.1021/nn2038884}
  {\bibfield  {journal} {\bibinfo  {journal} {ACS Nano}\ }\textbf {\bibinfo
  {volume} {6}},\ \bibinfo {pages} {1239--1250} (\bibinfo {year}
  {2012}{\natexlab{b}})}\BibitemShut {NoStop}%
\bibitem [{\citenamefont {Rabani}\ and\ \citenamefont
  {Baer}(2010)}]{Rabani2010227}%
  \BibitemOpen
  \bibfield  {author} {\bibinfo {author} {\bibfnamefont {Eran}\ \bibnamefont
  {Rabani}}\ and\ \bibinfo {author} {\bibfnamefont {Roi}\ \bibnamefont
  {Baer}},\ }\bibfield  {title} {\enquote {\bibinfo {title} {Theory of
  multiexciton generation in semiconductor nanocrystals},}\ }\href {\doibase
  10.1016/j.cplett.2010.07.059} {\bibfield  {journal} {\bibinfo  {journal}
  {Chemical Physics Letters}\ }\textbf {\bibinfo {volume} {496}},\ \bibinfo
  {pages} {227 -- 235} (\bibinfo {year} {2010})}\BibitemShut {NoStop}%
\bibitem [{\citenamefont {Jaeger}\ \emph {et~al.}(Article ASAP)\citenamefont
  {Jaeger}, \citenamefont {Hyeon-Deuk},\ and\ \citenamefont
  {Prezhdo}}]{doi:10.1021/ar3002365}%
  \BibitemOpen
  \bibfield  {author} {\bibinfo {author} {\bibfnamefont {Heather~M.}\
  \bibnamefont {Jaeger}}, \bibinfo {author} {\bibfnamefont {Kim}\ \bibnamefont
  {Hyeon-Deuk}}, \ and\ \bibinfo {author} {\bibfnamefont {Oleg~V.}\
  \bibnamefont {Prezhdo}},\ }\bibfield  {title} {\enquote {\bibinfo {title}
  {Exciton multiplication from first principles},}\ }\href {\doibase
  10.1021/ar3002365} {\bibfield  {journal} {\bibinfo  {journal} {Accounts of
  Chemical Research}\ } (\bibinfo {year} {Article ASAP}),\
  10.1021/ar3002365}\BibitemShut {NoStop}%
\bibitem [{\citenamefont {Vela}\ \emph {et~al.}(2010)\citenamefont {Vela},
  \citenamefont {Htoon}, \citenamefont {Chen}, \citenamefont {Park},
  \citenamefont {Ghosh}, \citenamefont {Goodwin}, \citenamefont {Werner},
  \citenamefont {Wells}, \citenamefont {Casson},\ and\ \citenamefont
  {Hollingsworth}}]{JBIO:JBIO201000058}%
  \BibitemOpen
  \bibfield  {author} {\bibinfo {author} {\bibfnamefont {Javier}\ \bibnamefont
  {Vela}}, \bibinfo {author} {\bibfnamefont {Han}\ \bibnamefont {Htoon}},
  \bibinfo {author} {\bibfnamefont {Yongfen}\ \bibnamefont {Chen}}, \bibinfo
  {author} {\bibfnamefont {Young-Shin}\ \bibnamefont {Park}}, \bibinfo {author}
  {\bibfnamefont {Yagnaseni}\ \bibnamefont {Ghosh}}, \bibinfo {author}
  {\bibfnamefont {Peter~M.}\ \bibnamefont {Goodwin}}, \bibinfo {author}
  {\bibfnamefont {James~H.}\ \bibnamefont {Werner}}, \bibinfo {author}
  {\bibfnamefont {Nathan~P.}\ \bibnamefont {Wells}}, \bibinfo {author}
  {\bibfnamefont {Joanna~L.}\ \bibnamefont {Casson}}, \ and\ \bibinfo {author}
  {\bibfnamefont {Jennifer~A.}\ \bibnamefont {Hollingsworth}},\ }\bibfield
  {title} {\enquote {\bibinfo {title} {Effect of shell thickness and
  composition on blinking suppression and the blinking mechanism in
  {``}giant{''} cdse/cds nanocrystal quantum dots},}\ }\href {\doibase
  10.1002/jbio.201000058} {\bibfield  {journal} {\bibinfo  {journal} {Journal
  of Biophotonics}\ }\textbf {\bibinfo {volume} {3}},\ \bibinfo {pages}
  {706--717} (\bibinfo {year} {2010})}\BibitemShut {NoStop}%
\bibitem [{\citenamefont {Rawalekar}\ \emph {et~al.}(2010)\citenamefont
  {Rawalekar}, \citenamefont {Kaniyankandy}, \citenamefont {Verma},\ and\
  \citenamefont {Ghosh}}]{doi:10.1021/jp909118c}%
  \BibitemOpen
  \bibfield  {author} {\bibinfo {author} {\bibfnamefont {Sachin}\ \bibnamefont
  {Rawalekar}}, \bibinfo {author} {\bibfnamefont {Sreejith}\ \bibnamefont
  {Kaniyankandy}}, \bibinfo {author} {\bibfnamefont {Sandeep}\ \bibnamefont
  {Verma}}, \ and\ \bibinfo {author} {\bibfnamefont {Hirendra~N.}\ \bibnamefont
  {Ghosh}},\ }\bibfield  {title} {\enquote {\bibinfo {title} {Ultrafast charge
  carrier relaxation and charge transfer dynamics of cdte/cds core-shell
  quantum dots as studied by femtosecond transient absorption spectroscopy},}\
  }\href {\doibase 10.1021/jp909118c} {\bibfield  {journal} {\bibinfo
  {journal} {The Journal of Physical Chemistry C}\ }\textbf {\bibinfo {volume}
  {114}},\ \bibinfo {pages} {1460--1466} (\bibinfo {year} {2010})}\BibitemShut
  {NoStop}%
\bibitem [{\citenamefont {Nemchinov}\ \emph {et~al.}(2008)\citenamefont
  {Nemchinov}, \citenamefont {Kirsanova}, \citenamefont {Hewa-Kasakarage},\
  and\ \citenamefont {Zamkov}}]{doi:10.1021/jp801523m}%
  \BibitemOpen
  \bibfield  {author} {\bibinfo {author} {\bibfnamefont {Alexander}\
  \bibnamefont {Nemchinov}}, \bibinfo {author} {\bibfnamefont {Maria}\
  \bibnamefont {Kirsanova}}, \bibinfo {author} {\bibfnamefont {Nishshanka~N.}\
  \bibnamefont {Hewa-Kasakarage}}, \ and\ \bibinfo {author} {\bibfnamefont
  {Mikhail}\ \bibnamefont {Zamkov}},\ }\bibfield  {title} {\enquote {\bibinfo
  {title} {Synthesis and characterization of type ii znse/cds core/shell
  nanocrystals},}\ }\href {\doibase 10.1021/jp801523m} {\bibfield  {journal}
  {\bibinfo  {journal} {The Journal of Physical Chemistry C}\ }\textbf
  {\bibinfo {volume} {112}},\ \bibinfo {pages} {9301--9307} (\bibinfo {year}
  {2008})}\BibitemShut {NoStop}%
\bibitem [{\citenamefont {Yan}\ \emph {et~al.}(2011)\citenamefont {Yan},
  \citenamefont {Chen},\ and\ \citenamefont
  {Van~Patten}}]{doi:10.1021/jp204420q}%
  \BibitemOpen
  \bibfield  {author} {\bibinfo {author} {\bibfnamefont {Yueran}\ \bibnamefont
  {Yan}}, \bibinfo {author} {\bibfnamefont {Gang}\ \bibnamefont {Chen}}, \ and\
  \bibinfo {author} {\bibfnamefont {P.~Gregory}\ \bibnamefont {Van~Patten}},\
  }\bibfield  {title} {\enquote {\bibinfo {title} {Ultrafast exciton dynamics
  in cdte nanocrystals and core/shell cdte/cds nanocrystals},}\ }\href
  {\doibase 10.1021/jp204420q} {\bibfield  {journal} {\bibinfo  {journal} {The
  Journal of Physical Chemistry C}\ }\textbf {\bibinfo {volume} {115}},\
  \bibinfo {pages} {22717--22728} (\bibinfo {year} {2011})}\BibitemShut
  {NoStop}%
\bibitem [{\citenamefont {Xu}\ \emph {et~al.}(2012)\citenamefont {Xu},
  \citenamefont {Hine}, \citenamefont {Maye}, \citenamefont {Meng},\ and\
  \citenamefont {Cotlet}}]{doi:10.1021/nn300525b}%
  \BibitemOpen
  \bibfield  {author} {\bibinfo {author} {\bibfnamefont {Zhihua}\ \bibnamefont
  {Xu}}, \bibinfo {author} {\bibfnamefont {Corey~R.}\ \bibnamefont {Hine}},
  \bibinfo {author} {\bibfnamefont {Mathew~M.}\ \bibnamefont {Maye}}, \bibinfo
  {author} {\bibfnamefont {Qingping}\ \bibnamefont {Meng}}, \ and\ \bibinfo
  {author} {\bibfnamefont {Mircea}\ \bibnamefont {Cotlet}},\ }\bibfield
  {title} {\enquote {\bibinfo {title} {Shell thickness dependent photoinduced
  hole transfer in hybrid conjugated polymer/quantum dot nanocomposites: From
  ensemble to single hybrid level},}\ }\href {\doibase 10.1021/nn300525b}
  {\bibfield  {journal} {\bibinfo  {journal} {ACS Nano}\ }\textbf {\bibinfo
  {volume} {6}},\ \bibinfo {pages} {4984--4992} (\bibinfo {year}
  {2012})}\BibitemShut {NoStop}%
\bibitem [{\citenamefont {Noguchi}\ \emph {et~al.}(2012)\citenamefont
  {Noguchi}, \citenamefont {Sugino}, \citenamefont {Nagaoka}, \citenamefont
  {Ishii},\ and\ \citenamefont {Ohno}}]{Noguchi2012}%
  \BibitemOpen
  \bibfield  {author} {\bibinfo {author} {\bibfnamefont {Y.}~\bibnamefont
  {Noguchi}}, \bibinfo {author} {\bibfnamefont {O.}~\bibnamefont {Sugino}},
  \bibinfo {author} {\bibfnamefont {M.}~\bibnamefont {Nagaoka}}, \bibinfo
  {author} {\bibfnamefont {S.}~\bibnamefont {Ishii}}, \ and\ \bibinfo {author}
  {\bibfnamefont {K.}~\bibnamefont {Ohno}},\ }\bibfield  {title} {\enquote
  {\bibinfo {title} {A gwbethe-salpeter calculation on photoabsorption spectra
  of (cdse) 3 and (cdse) 6 clusters},}\ }\href {\doibase 10.1063/1.4732123}
  {\bibfield  {journal} {\bibinfo  {journal} {Journal of Chemical Physics}\
  }\textbf {\bibinfo {volume} {137}},\ \bibinfo {pages} {024306} (\bibinfo
  {year} {2012})}\BibitemShut {NoStop}%
\bibitem [{\citenamefont {Lopez Del~Puerto}\ \emph {et~al.}(2008)\citenamefont
  {Lopez Del~Puerto}, \citenamefont {Tiago},\ and\ \citenamefont
  {Chelikowsky}}]{LopezDelPuerto2008}%
  \BibitemOpen
  \bibfield  {author} {\bibinfo {author} {\bibfnamefont {M.}~\bibnamefont
  {Lopez Del~Puerto}}, \bibinfo {author} {\bibfnamefont {M.L.}\ \bibnamefont
  {Tiago}}, \ and\ \bibinfo {author} {\bibfnamefont {J.R.}\ \bibnamefont
  {Chelikowsky}},\ }\bibfield  {title} {\enquote {\bibinfo {title} {Ab initio
  methods for the optical properties of cdse clusters},}\ }\href {\doibase
  10.1103/PhysRevB.77.045404} {\bibfield  {journal} {\bibinfo  {journal}
  {Physical Review B - Condensed Matter and Materials Physics}\ }\textbf
  {\bibinfo {volume} {77}},\ \bibinfo {pages} {045404} (\bibinfo {year}
  {2008})}\BibitemShut {NoStop}%
\bibitem [{\citenamefont {Del~Puerto}\ \emph {et~al.}(2006)\citenamefont
  {Del~Puerto}, \citenamefont {Tiago},\ and\ \citenamefont
  {Chelikowsky}}]{DelPuerto2006}%
  \BibitemOpen
  \bibfield  {author} {\bibinfo {author} {\bibfnamefont {M.L.}\ \bibnamefont
  {Del~Puerto}}, \bibinfo {author} {\bibfnamefont {M.L.}\ \bibnamefont
  {Tiago}}, \ and\ \bibinfo {author} {\bibfnamefont {J.R.}\ \bibnamefont
  {Chelikowsky}},\ }\bibfield  {title} {\enquote {\bibinfo {title} {Excitonic
  effects and optical properties of passivated cdse clusters},}\ }\href
  {\doibase 10.1103/PhysRevLett.97.096401} {\bibfield  {journal} {\bibinfo
  {journal} {Physical Review Letters}\ }\textbf {\bibinfo {volume} {97}},\
  \bibinfo {pages} {096401} (\bibinfo {year} {2006})}\BibitemShut {NoStop}%
\bibitem [{\citenamefont {Nguyen}\ \emph {et~al.}(2010)\citenamefont {Nguyen},
  \citenamefont {Day},\ and\ \citenamefont {Pachte}}]{Nguyen201016197}%
  \BibitemOpen
  \bibfield  {author} {\bibinfo {author} {\bibfnamefont {K.A.}\ \bibnamefont
  {Nguyen}}, \bibinfo {author} {\bibfnamefont {P.N.}\ \bibnamefont {Day}}, \
  and\ \bibinfo {author} {\bibnamefont {Pachte}},\ }\bibfield  {title}
  {\enquote {\bibinfo {title} {Understanding structural and optical properties
  of nanoscale cdse magic-size quantum dots: Insight from computational
  prediction},}\ }\href {\doibase 10.1021/jp103763d} {\bibfield  {journal}
  {\bibinfo  {journal} {Journal of Physical Chemistry C}\ }\textbf {\bibinfo
  {volume} {114}},\ \bibinfo {pages} {16197--16209} (\bibinfo {year}
  {2010})}\BibitemShut {NoStop}%
\bibitem [{\citenamefont {Yang}\ \emph {et~al.}(2011)\citenamefont {Yang},
  \citenamefont {Tretiak},\ and\ \citenamefont {Ivanov}}]{Yang2011405}%
  \BibitemOpen
  \bibfield  {author} {\bibinfo {author} {\bibfnamefont {P.}~\bibnamefont
  {Yang}}, \bibinfo {author} {\bibfnamefont {S.}~\bibnamefont {Tretiak}}, \
  and\ \bibinfo {author} {\bibfnamefont {S.}~\bibnamefont {Ivanov}},\
  }\bibfield  {title} {\enquote {\bibinfo {title} {Influence of surfactants and
  charges on cdse quantum dots},}\ }\href {\doibase 10.1007/s10876-011-0398-y}
  {\bibfield  {journal} {\bibinfo  {journal} {Journal of Cluster Science}\
  }\textbf {\bibinfo {volume} {22}},\ \bibinfo {pages} {405--431} (\bibinfo
  {year} {2011})}\BibitemShut {NoStop}%
\bibitem [{\citenamefont {Albert}\ \emph {et~al.}(2011)\citenamefont {Albert},
  \citenamefont {Ivanov}, \citenamefont {Tretiak},\ and\ \citenamefont
  {Kilina}}]{Albert201115793}%
  \BibitemOpen
  \bibfield  {author} {\bibinfo {author} {\bibfnamefont {V.V.}\ \bibnamefont
  {Albert}}, \bibinfo {author} {\bibfnamefont {S.A.}\ \bibnamefont {Ivanov}},
  \bibinfo {author} {\bibfnamefont {S.}~\bibnamefont {Tretiak}}, \ and\
  \bibinfo {author} {\bibfnamefont {S.V.}\ \bibnamefont {Kilina}},\ }\bibfield
  {title} {\enquote {\bibinfo {title} {Electronic structure of ligated cdse
  clusters: Dependence on dft methodology},}\ }\href {\doibase
  10.1021/jp202510z} {\bibfield  {journal} {\bibinfo  {journal} {Journal of
  Physical Chemistry C}\ }\textbf {\bibinfo {volume} {115}},\ \bibinfo {pages}
  {15793--15800} (\bibinfo {year} {2011})}\BibitemShut {NoStop}%
\bibitem [{\citenamefont {Kilin}\ \emph {et~al.}(2007)\citenamefont {Kilin},
  \citenamefont {Tsemekhman}, \citenamefont {Prezhdo}, \citenamefont
  {Zenkevich},\ and\ \citenamefont {von Borczyskowski}}]{Kilin2007342}%
  \BibitemOpen
  \bibfield  {author} {\bibinfo {author} {\bibfnamefont {D.S.}\ \bibnamefont
  {Kilin}}, \bibinfo {author} {\bibfnamefont {K.}~\bibnamefont {Tsemekhman}},
  \bibinfo {author} {\bibfnamefont {O.V.}\ \bibnamefont {Prezhdo}}, \bibinfo
  {author} {\bibfnamefont {E.I.}\ \bibnamefont {Zenkevich}}, \ and\ \bibinfo
  {author} {\bibfnamefont {C.}~\bibnamefont {von Borczyskowski}},\ }\bibfield
  {title} {\enquote {\bibinfo {title} {Ab initio study of exciton transfer
  dynamics from a core-shell semiconductor quantum dot to a
  porphyrin-sensitizer},}\ }\href {\doibase 10.1016/j.jphotochem.2007.02.017}
  {\bibfield  {journal} {\bibinfo  {journal} {Journal of Photochemistry and
  Photobiology A: Chemistry}\ }\textbf {\bibinfo {volume} {190}},\ \bibinfo
  {pages} {342--351} (\bibinfo {year} {2007})}\BibitemShut {NoStop}%
\bibitem [{\citenamefont {Liu}\ \emph {et~al.}(2009)\citenamefont {Liu},
  \citenamefont {Chung}, \citenamefont {Lee}, \citenamefont {Weiss},\ and\
  \citenamefont {Neuhauser}}]{Liu2009}%
  \BibitemOpen
  \bibfield  {author} {\bibinfo {author} {\bibfnamefont {C.}~\bibnamefont
  {Liu}}, \bibinfo {author} {\bibfnamefont {S.-Y.}\ \bibnamefont {Chung}},
  \bibinfo {author} {\bibfnamefont {S.}~\bibnamefont {Lee}}, \bibinfo {author}
  {\bibfnamefont {S.}~\bibnamefont {Weiss}}, \ and\ \bibinfo {author}
  {\bibfnamefont {D.}~\bibnamefont {Neuhauser}},\ }\bibfield  {title} {\enquote
  {\bibinfo {title} {Adsorbate-induced absorption redshift in an
  organic-inorganic cluster conjugate: Electronic effects of surfactants and
  organic adsorbates on the lowest excited states of a methanethiol-cdse
  conjugate},}\ }\href {\doibase 10.1063/1.3251774} {\bibfield  {journal}
  {\bibinfo  {journal} {Journal of Chemical Physics}\ }\textbf {\bibinfo
  {volume} {131}} (\bibinfo {year} {2009}),\ 10.1063/1.3251774}\BibitemShut
  {NoStop}%
\bibitem [{\citenamefont {Chung}\ \emph {et~al.}(2009)\citenamefont {Chung},
  \citenamefont {Lee}, \citenamefont {Liu},\ and\ \citenamefont
  {Neuhauser}}]{Chung2009292}%
  \BibitemOpen
  \bibfield  {author} {\bibinfo {author} {\bibfnamefont {S.-Y.}\ \bibnamefont
  {Chung}}, \bibinfo {author} {\bibfnamefont {S.}~\bibnamefont {Lee}}, \bibinfo
  {author} {\bibfnamefont {C.}~\bibnamefont {Liu}}, \ and\ \bibinfo {author}
  {\bibfnamefont {D.}~\bibnamefont {Neuhauser}},\ }\bibfield  {title} {\enquote
  {\bibinfo {title} {Structures and electronic spectra of cdse-cys complexes:
  Density functional theory study of a simple peptide-coated nanocluster},}\
  }\href {\doibase 10.1021/jp8062299} {\bibfield  {journal} {\bibinfo
  {journal} {Journal of Physical Chemistry B}\ }\textbf {\bibinfo {volume}
  {113}},\ \bibinfo {pages} {292--301} (\bibinfo {year} {2009})}\BibitemShut
  {NoStop}%
\bibitem [{\citenamefont {Kim}\ \emph {et~al.}(2010)\citenamefont {Kim},
  \citenamefont {Jang}, \citenamefont {Chung}, \citenamefont {Lee},
  \citenamefont {Lee}, \citenamefont {Kim}, \citenamefont {Liu},\ and\
  \citenamefont {Neuhauser}}]{Kim2010471}%
  \BibitemOpen
  \bibfield  {author} {\bibinfo {author} {\bibfnamefont {H.S.}\ \bibnamefont
  {Kim}}, \bibinfo {author} {\bibfnamefont {S.-W.}\ \bibnamefont {Jang}},
  \bibinfo {author} {\bibfnamefont {S.Y.}\ \bibnamefont {Chung}}, \bibinfo
  {author} {\bibfnamefont {S.}~\bibnamefont {Lee}}, \bibinfo {author}
  {\bibfnamefont {Y.}~\bibnamefont {Lee}}, \bibinfo {author} {\bibfnamefont
  {B.}~\bibnamefont {Kim}}, \bibinfo {author} {\bibfnamefont {C.}~\bibnamefont
  {Liu}}, \ and\ \bibinfo {author} {\bibfnamefont {D.}~\bibnamefont
  {Neuhauser}},\ }\bibfield  {title} {\enquote {\bibinfo {title} {Effects of
  bioconjugation on the structures and electronic spectra of cdse: Density
  functional theory study of cdse - adenine complexes},}\ }\href {\doibase
  10.1021/jp907725f} {\bibfield  {journal} {\bibinfo  {journal} {Journal of
  Physical Chemistry B}\ }\textbf {\bibinfo {volume} {114}},\ \bibinfo {pages}
  {471--479} (\bibinfo {year} {2010})}\BibitemShut {NoStop}%
\bibitem [{\citenamefont {Nadler}\ and\ \citenamefont
  {Sanz}(2013)}]{Nadler20131}%
  \BibitemOpen
  \bibfield  {author} {\bibinfo {author} {\bibfnamefont {R.}~\bibnamefont
  {Nadler}}\ and\ \bibinfo {author} {\bibfnamefont {J.F.}\ \bibnamefont
  {Sanz}},\ }\bibfield  {title} {\enquote {\bibinfo {title} {Simulating the
  optical properties of cdse clusters using the rt-tddft approach},}\ }\href
  {\doibase 10.1007/s00214-013-1342-z} {\bibfield  {journal} {\bibinfo
  {journal} {Theoretical Chemistry Accounts}\ }\textbf {\bibinfo {volume}
  {132}},\ \bibinfo {pages} {1--9} (\bibinfo {year} {2013})}\BibitemShut
  {NoStop}%
\bibitem [{\citenamefont {Abuelela}\ \emph {et~al.}(2012)\citenamefont
  {Abuelela}, \citenamefont {Mohamed},\ and\ \citenamefont
  {Prezhdo}}]{Abuelela201214674}%
  \BibitemOpen
  \bibfield  {author} {\bibinfo {author} {\bibfnamefont {A.M.}\ \bibnamefont
  {Abuelela}}, \bibinfo {author} {\bibfnamefont {T.A.}\ \bibnamefont
  {Mohamed}}, \ and\ \bibinfo {author} {\bibfnamefont {O.V.}\ \bibnamefont
  {Prezhdo}},\ }\bibfield  {title} {\enquote {\bibinfo {title} {Dft simulation
  and vibrational analysis of the ir and raman spectra of a cdse quantum dot
  capped by methylamine and trimethylphosphine oxide ligands},}\ }\href
  {\doibase 10.1021/jp303275v} {\bibfield  {journal} {\bibinfo  {journal}
  {Journal of Physical Chemistry C}\ }\textbf {\bibinfo {volume} {116}},\
  \bibinfo {pages} {14674--14681} (\bibinfo {year} {2012})}\BibitemShut
  {NoStop}%
\bibitem [{\citenamefont {Fischer}\ \emph {et~al.}(2012)\citenamefont
  {Fischer}, \citenamefont {Crotty}, \citenamefont {Kilina}, \citenamefont
  {Ivanov},\ and\ \citenamefont {Tretiak}}]{Fischer2012904}%
  \BibitemOpen
  \bibfield  {author} {\bibinfo {author} {\bibfnamefont {S.A.}\ \bibnamefont
  {Fischer}}, \bibinfo {author} {\bibfnamefont {A.M.}\ \bibnamefont {Crotty}},
  \bibinfo {author} {\bibfnamefont {S.V.}\ \bibnamefont {Kilina}}, \bibinfo
  {author} {\bibfnamefont {S.A.}\ \bibnamefont {Ivanov}}, \ and\ \bibinfo
  {author} {\bibfnamefont {S.}~\bibnamefont {Tretiak}},\ }\bibfield  {title}
  {\enquote {\bibinfo {title} {Passivating ligand and solvent contributions to
  the electronic properties of semiconductor nanocrystals},}\ }\href {\doibase
  10.1039/c2nr11398h} {\bibfield  {journal} {\bibinfo  {journal} {Nanoscale}\
  }\textbf {\bibinfo {volume} {4}},\ \bibinfo {pages} {904--914} (\bibinfo
  {year} {2012})}\BibitemShut {NoStop}%
\bibitem [{\citenamefont {Del~Ben}\ \emph {et~al.}(2011)\citenamefont
  {Del~Ben}, \citenamefont {Havenith}, \citenamefont {Broer},\ and\
  \citenamefont {Stener}}]{DelBen201116782}%
  \BibitemOpen
  \bibfield  {author} {\bibinfo {author} {\bibfnamefont {M.}~\bibnamefont
  {Del~Ben}}, \bibinfo {author} {\bibfnamefont {R.W.A.}\ \bibnamefont
  {Havenith}}, \bibinfo {author} {\bibfnamefont {R.}~\bibnamefont {Broer}}, \
  and\ \bibinfo {author} {\bibfnamefont {M.}~\bibnamefont {Stener}},\
  }\bibfield  {title} {\enquote {\bibinfo {title} {Density functional study on
  the morphology and photoabsorption of cdse nanoclusters},}\ }\href {\doibase
  10.1021/jp203686a} {\bibfield  {journal} {\bibinfo  {journal} {Journal of
  Physical Chemistry C}\ }\textbf {\bibinfo {volume} {115}},\ \bibinfo {pages}
  {16782--16796} (\bibinfo {year} {2011})}\BibitemShut {NoStop}%
\bibitem [{\citenamefont {Turkowski}\ \emph {et~al.}(2009)\citenamefont
  {Turkowski}, \citenamefont {Leonardo},\ and\ \citenamefont
  {Ullrich}}]{Turkowski2009}%
  \BibitemOpen
  \bibfield  {author} {\bibinfo {author} {\bibfnamefont {V.}~\bibnamefont
  {Turkowski}}, \bibinfo {author} {\bibfnamefont {A.}~\bibnamefont {Leonardo}},
  \ and\ \bibinfo {author} {\bibfnamefont {C.A.}\ \bibnamefont {Ullrich}},\
  }\bibfield  {title} {\enquote {\bibinfo {title} {Time-dependent
  density-functional approach for exciton binding energies},}\ }\href {\doibase
  10.1103/PhysRevB.79.233201} {\bibfield  {journal} {\bibinfo  {journal}
  {Physical Review B - Condensed Matter and Materials Physics}\ }\textbf
  {\bibinfo {volume} {79}} (\bibinfo {year} {2009}),\
  10.1103/PhysRevB.79.233201}\BibitemShut {NoStop}%
\bibitem [{\citenamefont {Li}\ and\ \citenamefont {Ullrich}(2011)}]{Li2011157}%
  \BibitemOpen
  \bibfield  {author} {\bibinfo {author} {\bibfnamefont {Y.}~\bibnamefont
  {Li}}\ and\ \bibinfo {author} {\bibfnamefont {C.A.}\ \bibnamefont
  {Ullrich}},\ }\bibfield  {title} {\enquote {\bibinfo {title} {Time-dependent
  transition density matrix},}\ }\href {\doibase
  10.1016/j.chemphys.2011.02.001} {\bibfield  {journal} {\bibinfo  {journal}
  {Chemical Physics}\ }\textbf {\bibinfo {volume} {391}},\ \bibinfo {pages}
  {157--163} (\bibinfo {year} {2011})}\BibitemShut {NoStop}%
\bibitem [{\citenamefont {Yang}\ \emph {et~al.}(2012)\citenamefont {Yang},
  \citenamefont {Li},\ and\ \citenamefont {Ullrich}}]{Yang2012}%
  \BibitemOpen
  \bibfield  {author} {\bibinfo {author} {\bibfnamefont {Z.-H.}\ \bibnamefont
  {Yang}}, \bibinfo {author} {\bibfnamefont {Y.}~\bibnamefont {Li}}, \ and\
  \bibinfo {author} {\bibfnamefont {C.A.}\ \bibnamefont {Ullrich}},\ }\bibfield
   {title} {\enquote {\bibinfo {title} {A minimal model for excitons within
  time-dependent density-functional theory},}\ }\href {\doibase
  10.1063/1.4730031} {\bibfield  {journal} {\bibinfo  {journal} {Journal of
  Chemical Physics}\ }\textbf {\bibinfo {volume} {137}} (\bibinfo {year}
  {2012}),\ 10.1063/1.4730031}\BibitemShut {NoStop}%
\bibitem [{\citenamefont {Yang}\ and\ \citenamefont
  {Ullrich}(2013)}]{Yang2013}%
  \BibitemOpen
  \bibfield  {author} {\bibinfo {author} {\bibfnamefont {Z.-H.}\ \bibnamefont
  {Yang}}\ and\ \bibinfo {author} {\bibfnamefont {C.A.}\ \bibnamefont
  {Ullrich}},\ }\bibfield  {title} {\enquote {\bibinfo {title} {Direct
  calculation of exciton binding energies with time-dependent
  density-functional theory},}\ }\href {\doibase 10.1103/PhysRevB.87.195204}
  {\bibfield  {journal} {\bibinfo  {journal} {Physical Review B - Condensed
  Matter and Materials Physics}\ }\textbf {\bibinfo {volume} {87}} (\bibinfo
  {year} {2013}),\ 10.1103/PhysRevB.87.195204}\BibitemShut {NoStop}%
\bibitem [{\citenamefont {Neuhauser}\ \emph {et~al.}(2013)\citenamefont
  {Neuhauser}, \citenamefont {Rabani},\ and\ \citenamefont
  {Baer}}]{Neuhauser20131172}%
  \BibitemOpen
  \bibfield  {author} {\bibinfo {author} {\bibfnamefont {D.}~\bibnamefont
  {Neuhauser}}, \bibinfo {author} {\bibfnamefont {E.}~\bibnamefont {Rabani}}, \
  and\ \bibinfo {author} {\bibfnamefont {R.}~\bibnamefont {Baer}},\ }\bibfield
  {title} {\enquote {\bibinfo {title} {Expeditious stochastic calculation of
  random-phase approximation energies for thousands of electrons in three
  dimensions},}\ }\href {\doibase 10.1021/jz3021606} {\bibfield  {journal}
  {\bibinfo  {journal} {Journal of Physical Chemistry Letters}\ }\textbf
  {\bibinfo {volume} {4}},\ \bibinfo {pages} {1172--1176} (\bibinfo {year}
  {2013})}\BibitemShut {NoStop}%
\bibitem [{\citenamefont {Franceschetti}\ \emph {et~al.}(1999)\citenamefont
  {Franceschetti}, \citenamefont {Fu}, \citenamefont {Wang},\ and\
  \citenamefont {Zunger}}]{PhysRevB.60.1819}%
  \BibitemOpen
  \bibfield  {author} {\bibinfo {author} {\bibfnamefont {A.}~\bibnamefont
  {Franceschetti}}, \bibinfo {author} {\bibfnamefont {H.}~\bibnamefont {Fu}},
  \bibinfo {author} {\bibfnamefont {L.~W.}\ \bibnamefont {Wang}}, \ and\
  \bibinfo {author} {\bibfnamefont {A.}~\bibnamefont {Zunger}},\ }\bibfield
  {title} {\enquote {\bibinfo {title} {Many-body pseudopotential theory of
  excitons in inp and cdse quantum dots},}\ }\href {\doibase
  10.1103/PhysRevB.60.1819} {\bibfield  {journal} {\bibinfo  {journal} {Phys.
  Rev. B}\ }\textbf {\bibinfo {volume} {60}},\ \bibinfo {pages} {1819--1829}
  (\bibinfo {year} {1999})}\BibitemShut {NoStop}%
\bibitem [{\citenamefont {Baer}\ and\ \citenamefont
  {Rabani}(2013)}]{baer:051102}%
  \BibitemOpen
  \bibfield  {author} {\bibinfo {author} {\bibfnamefont {Roi}\ \bibnamefont
  {Baer}}\ and\ \bibinfo {author} {\bibfnamefont {Eran}\ \bibnamefont
  {Rabani}},\ }\bibfield  {title} {\enquote {\bibinfo {title} {Communication:
  Biexciton generation rates in cdse nanorods are length independent},}\ }\href
  {\doibase 10.1063/1.4790600} {\bibfield  {journal} {\bibinfo  {journal} {The
  Journal of Chemical Physics}\ }\textbf {\bibinfo {volume} {138}},\ \bibinfo
  {eid} {051102} (\bibinfo {year} {2013})}\BibitemShut {NoStop}%
\bibitem [{\citenamefont {Rabani}\ \emph {et~al.}(1999)\citenamefont {Rabani},
  \citenamefont {Hetenyi}, \citenamefont {Berne},\ and\ \citenamefont
  {Brus}}]{rabani:5355}%
  \BibitemOpen
  \bibfield  {author} {\bibinfo {author} {\bibfnamefont {Eran}\ \bibnamefont
  {Rabani}}, \bibinfo {author} {\bibfnamefont {Balazs}\ \bibnamefont
  {Hetenyi}}, \bibinfo {author} {\bibfnamefont {B.~J.}\ \bibnamefont {Berne}},
  \ and\ \bibinfo {author} {\bibfnamefont {L.~E.}\ \bibnamefont {Brus}},\
  }\bibfield  {title} {\enquote {\bibinfo {title} {Electronic properties of
  cdse nanocrystals in the absence and presence of a dielectric medium},}\
  }\href {\doibase 10.1063/1.478431} {\bibfield  {journal} {\bibinfo  {journal}
  {The Journal of Chemical Physics}\ }\textbf {\bibinfo {volume} {110}},\
  \bibinfo {pages} {5355--5369} (\bibinfo {year} {1999})}\BibitemShut {NoStop}%
\bibitem [{\citenamefont {Wang}\ and\ \citenamefont
  {Zunger}(1994)}]{Wang19942394}%
  \BibitemOpen
  \bibfield  {author} {\bibinfo {author} {\bibfnamefont {L.-W.}\ \bibnamefont
  {Wang}}\ and\ \bibinfo {author} {\bibfnamefont {A.}~\bibnamefont {Zunger}},\
  }\bibfield  {title} {\enquote {\bibinfo {title} {Solving schr\"{o}dinger's
  equation around a desired energy: Application to silicon quantum dots},}\
  }\href {\doibase 10.1063/1.466486} {\bibfield  {journal} {\bibinfo  {journal}
  {The Journal of Chemical Physics}\ }\textbf {\bibinfo {volume} {100}},\
  \bibinfo {pages} {2394--2397} (\bibinfo {year} {1994})}\BibitemShut {NoStop}%
\bibitem [{\citenamefont {Canning}\ \emph {et~al.}(2000)\citenamefont
  {Canning}, \citenamefont {Wang}, \citenamefont {Williamson},\ and\
  \citenamefont {Zunger}}]{Canning200029}%
  \BibitemOpen
  \bibfield  {author} {\bibinfo {author} {\bibfnamefont {A.}~\bibnamefont
  {Canning}}, \bibinfo {author} {\bibfnamefont {L.W.}\ \bibnamefont {Wang}},
  \bibinfo {author} {\bibfnamefont {A.}~\bibnamefont {Williamson}}, \ and\
  \bibinfo {author} {\bibfnamefont {A.}~\bibnamefont {Zunger}},\ }\bibfield
  {title} {\enquote {\bibinfo {title} {Parallel empirical pseudopotential
  electronic structure calculations for million atom systems},}\ }\href
  {\doibase 10.1006/jcph.2000.6440} {\bibfield  {journal} {\bibinfo  {journal}
  {Journal of Computational Physics}\ }\textbf {\bibinfo {volume} {160}},\
  \bibinfo {pages} {29--41} (\bibinfo {year} {2000})}\BibitemShut {NoStop}%
\bibitem [{\citenamefont {Toledo}\ and\ \citenamefont
  {Rabani}(2002)}]{Toledo2002256}%
  \BibitemOpen
  \bibfield  {author} {\bibinfo {author} {\bibfnamefont {S.}~\bibnamefont
  {Toledo}}\ and\ \bibinfo {author} {\bibfnamefont {E.}~\bibnamefont
  {Rabani}},\ }\bibfield  {title} {\enquote {\bibinfo {title} {Very large
  electronic structure calculations using an out-of-core filter-diagonalization
  method},}\ }\href {\doibase 10.1006/jcph.2002.7090} {\bibfield  {journal}
  {\bibinfo  {journal} {Journal of Computational Physics}\ }\textbf {\bibinfo
  {volume} {180}},\ \bibinfo {pages} {256--269} (\bibinfo {year}
  {2002})}\BibitemShut {NoStop}%
\bibitem [{\citenamefont {V\"{o}mel}\ \emph {et~al.}(2008)\citenamefont
  {V\"{o}mel}, \citenamefont {Tomov}, \citenamefont {Marques}, \citenamefont
  {Canning}, \citenamefont {Wang},\ and\ \citenamefont
  {Dongarra}}]{Vomel20087113}%
  \BibitemOpen
  \bibfield  {author} {\bibinfo {author} {\bibfnamefont {C.}~\bibnamefont
  {V\"{o}mel}}, \bibinfo {author} {\bibfnamefont {S.Z.}\ \bibnamefont {Tomov}},
  \bibinfo {author} {\bibfnamefont {O.A.}\ \bibnamefont {Marques}}, \bibinfo
  {author} {\bibfnamefont {A.}~\bibnamefont {Canning}}, \bibinfo {author}
  {\bibfnamefont {L.-W.}\ \bibnamefont {Wang}}, \ and\ \bibinfo {author}
  {\bibfnamefont {J.J.}\ \bibnamefont {Dongarra}},\ }\bibfield  {title}
  {\enquote {\bibinfo {title} {State-of-the-art eigensolvers for electronic
  structure calculations of large scale nano-systems},}\ }\href {\doibase
  10.1016/j.jcp.2008.01.018} {\bibfield  {journal} {\bibinfo  {journal}
  {Journal of Computational Physics}\ }\textbf {\bibinfo {volume} {227}},\
  \bibinfo {pages} {7113--7124} (\bibinfo {year} {2008})}\BibitemShut {NoStop}%
\bibitem [{\citenamefont {Jordan}\ \emph {et~al.}(2012)\citenamefont {Jordan},
  \citenamefont {Marsman}, \citenamefont {Kim},\ and\ \citenamefont
  {Kresse}}]{Jordan20124836}%
  \BibitemOpen
  \bibfield  {author} {\bibinfo {author} {\bibfnamefont {G.}~\bibnamefont
  {Jordan}}, \bibinfo {author} {\bibfnamefont {M.}~\bibnamefont {Marsman}},
  \bibinfo {author} {\bibfnamefont {Y.-S.}\ \bibnamefont {Kim}}, \ and\
  \bibinfo {author} {\bibfnamefont {G.}~\bibnamefont {Kresse}},\ }\bibfield
  {title} {\enquote {\bibinfo {title} {Fast iterative interior eigensolver for
  millions of atoms},}\ }\href {\doibase 10.1016/j.jcp.2012.04.010} {\bibfield
  {journal} {\bibinfo  {journal} {Journal of Computational Physics}\ }\textbf
  {\bibinfo {volume} {231}},\ \bibinfo {pages} {4836--4847} (\bibinfo {year}
  {2012})}\BibitemShut {NoStop}%
\bibitem [{\citenamefont {Mazziotti}(2007)}]{mazziotti2007advances}%
  \BibitemOpen
  \bibfield  {author} {\bibinfo {author} {\bibfnamefont {D.A.}\ \bibnamefont
  {Mazziotti}},\ }\href {http://books.google.com/books?id=\_gvsahY8sjoC} {\emph
  {\bibinfo {title} {Advances in Chemical Physics, Reduced-Density-Matrix
  Mechanics: With Application to Many-Electron Atoms and Molecules}}},\
  Advances in Chemical Physics\ (\bibinfo  {publisher} {Wiley},\ \bibinfo
  {year} {2007})\BibitemShut {NoStop}%
\bibitem [{\citenamefont {Elward}\ \emph
  {et~al.}(2012{\natexlab{a}})\citenamefont {Elward}, \citenamefont
  {Thallinger},\ and\ \citenamefont {Chakraborty}}]{elward:124105}%
  \BibitemOpen
  \bibfield  {author} {\bibinfo {author} {\bibfnamefont {Jennifer~M.}\
  \bibnamefont {Elward}}, \bibinfo {author} {\bibfnamefont {Barbara}\
  \bibnamefont {Thallinger}}, \ and\ \bibinfo {author} {\bibfnamefont
  {Arindam}\ \bibnamefont {Chakraborty}},\ }\bibfield  {title} {\enquote
  {\bibinfo {title} {Calculation of electron-hole recombination probability
  using explicitly correlated hartree-fock method},}\ }\href {\doibase
  10.1063/1.3693765} {\bibfield  {journal} {\bibinfo  {journal} {The Journal of
  Chemical Physics}\ }\textbf {\bibinfo {volume} {136}},\ \bibinfo {eid}
  {124105} (\bibinfo {year} {2012}{\natexlab{a}})}\BibitemShut {NoStop}%
\bibitem [{\citenamefont {Elward}\ \emph
  {et~al.}(2012{\natexlab{b}})\citenamefont {Elward}, \citenamefont {Hoffman},\
  and\ \citenamefont {Chakraborty}}]{Elward2012182}%
  \BibitemOpen
  \bibfield  {author} {\bibinfo {author} {\bibfnamefont {Jennifer~M.}\
  \bibnamefont {Elward}}, \bibinfo {author} {\bibfnamefont {Jacob}\
  \bibnamefont {Hoffman}}, \ and\ \bibinfo {author} {\bibfnamefont {Arindam}\
  \bibnamefont {Chakraborty}},\ }\bibfield  {title} {\enquote {\bibinfo {title}
  {Investigation of electron-hole correlation using explicitly correlated
  configuration interaction method},}\ }\href {\doibase
  10.1016/j.cplett.2012.03.050} {\bibfield  {journal} {\bibinfo  {journal}
  {Chemical Physics Letters}\ }\textbf {\bibinfo {volume} {535}},\ \bibinfo
  {pages} {182 -- 186} (\bibinfo {year} {2012}{\natexlab{b}})}\BibitemShut
  {NoStop}%
\bibitem [{\citenamefont {Wimmer}\ \emph
  {et~al.}(2006{\natexlab{a}})\citenamefont {Wimmer}, \citenamefont {Nair},\
  and\ \citenamefont {Shumway}}]{RefWorks:23}%
  \BibitemOpen
  \bibfield  {author} {\bibinfo {author} {\bibfnamefont {M.}~\bibnamefont
  {Wimmer}}, \bibinfo {author} {\bibfnamefont {S.~V.}\ \bibnamefont {Nair}}, \
  and\ \bibinfo {author} {\bibfnamefont {J.}~\bibnamefont {Shumway}},\
  }\bibfield  {title} {\enquote {\bibinfo {title} {Biexciton recombination
  rates in self-assembled quantum dots},}\ }\href {\doibase
  10.1103/PhysRevB.73.165305} {\bibfield  {journal} {\bibinfo  {journal}
  {Physical Review B - Condensed Matter and Materials Physics}\ }\textbf
  {\bibinfo {volume} {73}},\ \bibinfo {pages} {1--10} (\bibinfo {year}
  {2006}{\natexlab{a}})}\BibitemShut {NoStop}%
\bibitem [{\citenamefont {Cancio}\ and\ \citenamefont
  {Chang}(1990)}]{Cancio199011317}%
  \BibitemOpen
  \bibfield  {author} {\bibinfo {author} {\bibfnamefont {A.C.}\ \bibnamefont
  {Cancio}}\ and\ \bibinfo {author} {\bibfnamefont {Y.-C.}\ \bibnamefont
  {Chang}},\ }\bibfield  {title} {\enquote {\bibinfo {title} {Quantum monte
  carlo study of polyexcitons in semiconductors},}\ }\href {\doibase
  10.1103/PhysRevB.42.11317} {\bibfield  {journal} {\bibinfo  {journal}
  {Physical Review B}\ }\textbf {\bibinfo {volume} {42}},\ \bibinfo {pages}
  {11317--11324} (\bibinfo {year} {1990})}\BibitemShut {NoStop}%
\bibitem [{\citenamefont {Cancio}\ and\ \citenamefont
  {Chang}(1993)}]{Cancio199313246}%
  \BibitemOpen
  \bibfield  {author} {\bibinfo {author} {\bibfnamefont {A.C.}\ \bibnamefont
  {Cancio}}\ and\ \bibinfo {author} {\bibfnamefont {Y.-C.}\ \bibnamefont
  {Chang}},\ }\bibfield  {title} {\enquote {\bibinfo {title} {Quantum monte
  carlo studies of binding energy and radiative lifetime of bound excitons in
  direct-gap semiconductors},}\ }\href {\doibase 10.1103/PhysRevB.47.13246}
  {\bibfield  {journal} {\bibinfo  {journal} {Physical Review B}\ }\textbf
  {\bibinfo {volume} {47}},\ \bibinfo {pages} {13246--13259} (\bibinfo {year}
  {1993})}\BibitemShut {NoStop}%
\bibitem [{\citenamefont {Zhu}\ \emph {et~al.}(1996)\citenamefont {Zhu},
  \citenamefont {Hybertsen},\ and\ \citenamefont
  {Littlewood}}]{PhysRevB.54.13575}%
  \BibitemOpen
  \bibfield  {author} {\bibinfo {author} {\bibfnamefont {Xuejun}\ \bibnamefont
  {Zhu}}, \bibinfo {author} {\bibfnamefont {Mark~S.}\ \bibnamefont
  {Hybertsen}}, \ and\ \bibinfo {author} {\bibfnamefont {P.~B.}\ \bibnamefont
  {Littlewood}},\ }\bibfield  {title} {\enquote {\bibinfo {title}
  {Electron-hole system revisited: A variational quantum monte carlo study},}\
  }\href {\doibase 10.1103/PhysRevB.54.13575} {\bibfield  {journal} {\bibinfo
  {journal} {Phys. Rev. B}\ }\textbf {\bibinfo {volume} {54}},\ \bibinfo
  {pages} {13575--13580} (\bibinfo {year} {1996})}\BibitemShut {NoStop}%
\bibitem [{\citenamefont {Boys}(1960)}]{Boys25101960}%
  \BibitemOpen
  \bibfield  {author} {\bibinfo {author} {\bibfnamefont {S.~F.}\ \bibnamefont
  {Boys}},\ }\bibfield  {title} {\enquote {\bibinfo {title} {The integral
  formulae for the variational solution of the molecular many-electron wave
  equations in terms of gaussian functions with direct electronic
  correlation},}\ }\href {\doibase 10.1098/rspa.1960.0195} {\bibfield
  {journal} {\bibinfo  {journal} {Proceedings of the Royal Society of London.
  Series A. Mathematical and Physical Sciences}\ }\textbf {\bibinfo {volume}
  {258}},\ \bibinfo {pages} {402--411} (\bibinfo {year} {1960})}\BibitemShut
  {NoStop}%
\bibitem [{\citenamefont {Persson}\ and\ \citenamefont
  {Taylor}(1996)}]{persson:5915}%
  \BibitemOpen
  \bibfield  {author} {\bibinfo {author} {\bibfnamefont {B.~Joakim}\
  \bibnamefont {Persson}}\ and\ \bibinfo {author} {\bibfnamefont {Peter~R.}\
  \bibnamefont {Taylor}},\ }\bibfield  {title} {\enquote {\bibinfo {title}
  {Accurate quantum-chemical calculations: The use of gaussian-type geminal
  functions in the treatment of electron correlation},}\ }\href {\doibase
  10.1063/1.472432} {\bibfield  {journal} {\bibinfo  {journal} {The Journal of
  Chemical Physics}\ }\textbf {\bibinfo {volume} {105}},\ \bibinfo {pages}
  {5915--5926} (\bibinfo {year} {1996})}\BibitemShut {NoStop}%
\bibitem [{\citenamefont {Hu}\ \emph {et~al.}(1990)\citenamefont {Hu},
  \citenamefont {Lindberg},\ and\ \citenamefont {Koch}}]{PhysRevB.42.1713}%
  \BibitemOpen
  \bibfield  {author} {\bibinfo {author} {\bibfnamefont {Y.~Z.}\ \bibnamefont
  {Hu}}, \bibinfo {author} {\bibfnamefont {M.}~\bibnamefont {Lindberg}}, \ and\
  \bibinfo {author} {\bibfnamefont {S.~W.}\ \bibnamefont {Koch}},\ }\bibfield
  {title} {\enquote {\bibinfo {title} {Theory of optically excited intrinsic
  semiconductor quantum dots},}\ }\href {\doibase 10.1103/PhysRevB.42.1713}
  {\bibfield  {journal} {\bibinfo  {journal} {Phys. Rev. B}\ }\textbf {\bibinfo
  {volume} {42}},\ \bibinfo {pages} {1713--1723} (\bibinfo {year}
  {1990})}\BibitemShut {NoStop}%
\bibitem [{\citenamefont {Burovski}\ \emph {et~al.}(2001)\citenamefont
  {Burovski}, \citenamefont {Mishchenko}, \citenamefont {Prokof\'{e}v},\ and\
  \citenamefont {Svistunov}}]{PhysRevLett.87.186402}%
  \BibitemOpen
  \bibfield  {author} {\bibinfo {author} {\bibfnamefont {E.~A.}\ \bibnamefont
  {Burovski}}, \bibinfo {author} {\bibfnamefont {A.~S.}\ \bibnamefont
  {Mishchenko}}, \bibinfo {author} {\bibfnamefont {N.~V.}\ \bibnamefont
  {Prokof\'{e}v}}, \ and\ \bibinfo {author} {\bibfnamefont {B.~V.}\
  \bibnamefont {Svistunov}},\ }\bibfield  {title} {\enquote {\bibinfo {title}
  {Diagrammatic quantum monte carlo for two-body problems: Applied to
  excitons},}\ }\href {\doibase 10.1103/PhysRevLett.87.186402} {\bibfield
  {journal} {\bibinfo  {journal} {Phys. Rev. Lett.}\ }\textbf {\bibinfo
  {volume} {87}},\ \bibinfo {pages} {186402} (\bibinfo {year}
  {2001})}\BibitemShut {NoStop}%
\bibitem [{\citenamefont {Wimmer}\ \emph
  {et~al.}(2006{\natexlab{b}})\citenamefont {Wimmer}, \citenamefont {Nair},\
  and\ \citenamefont {Shumway}}]{PhysRevB.73.165305}%
  \BibitemOpen
  \bibfield  {author} {\bibinfo {author} {\bibfnamefont {Michael}\ \bibnamefont
  {Wimmer}}, \bibinfo {author} {\bibfnamefont {S.~V.}\ \bibnamefont {Nair}}, \
  and\ \bibinfo {author} {\bibfnamefont {J.}~\bibnamefont {Shumway}},\
  }\bibfield  {title} {\enquote {\bibinfo {title} {Biexciton recombination
  rates in self-assembled quantum dots},}\ }\href {\doibase
  10.1103/PhysRevB.73.165305} {\bibfield  {journal} {\bibinfo  {journal} {Phys.
  Rev. B}\ }\textbf {\bibinfo {volume} {73}},\ \bibinfo {pages} {165305}
  (\bibinfo {year} {2006}{\natexlab{b}})}\BibitemShut {NoStop}%
\bibitem [{\citenamefont {Woggon}(1997)}]{woggon1997springer}%
  \BibitemOpen
  \bibfield  {author} {\bibinfo {author} {\bibfnamefont {U.}~\bibnamefont
  {Woggon}},\ }\href@noop {} {\emph {\bibinfo {title} {Springer Tracts in
  Modern Physics}}},\ \bibinfo {series} {Springer Tracts in Modern Physics}\
  No.\ \bibinfo {number} {v. 136}\ (\bibinfo  {publisher} {Springer-Verlag.},\
  \bibinfo {year} {1997})\BibitemShut {NoStop}%
\bibitem [{\citenamefont {Brask\'{e}n}\ \emph {et~al.}(2001)\citenamefont
  {Brask\'{e}n}, \citenamefont {Lindberg}, \citenamefont {Sundholm},\ and\
  \citenamefont {Olsen}}]{Brasken2001775}%
  \BibitemOpen
  \bibfield  {author} {\bibinfo {author} {\bibfnamefont {M.}~\bibnamefont
  {Brask\'{e}n}}, \bibinfo {author} {\bibfnamefont {M.}~\bibnamefont
  {Lindberg}}, \bibinfo {author} {\bibfnamefont {D.}~\bibnamefont {Sundholm}},
  \ and\ \bibinfo {author} {\bibfnamefont {J.}~\bibnamefont {Olsen}},\
  }\bibfield  {title} {\enquote {\bibinfo {title} {Full configuration
  interaction calculations of electron-hole correlation effects in
  strain-induced quantum dots},}\ }\href {\doibase 10.1103/PhysRevB.61.7652}
  {\bibfield  {journal} {\bibinfo  {journal} {Physica Status Solidi (B) Basic
  Research}\ }\textbf {\bibinfo {volume} {224}},\ \bibinfo {pages} {775--779}
  (\bibinfo {year} {2001})}\BibitemShut {NoStop}%
\bibitem [{\citenamefont {Corni}\ \emph
  {et~al.}(2003{\natexlab{a}})\citenamefont {Corni}, \citenamefont
  {Brask\'{e}n}, \citenamefont {Lindberg}, \citenamefont {Olsen},\ and\
  \citenamefont {Sundholm}}]{Corni2003436}%
  \BibitemOpen
  \bibfield  {author} {\bibinfo {author} {\bibfnamefont {S.}~\bibnamefont
  {Corni}}, \bibinfo {author} {\bibfnamefont {M.}~\bibnamefont {Brask\'{e}n}},
  \bibinfo {author} {\bibfnamefont {M.}~\bibnamefont {Lindberg}}, \bibinfo
  {author} {\bibfnamefont {J.}~\bibnamefont {Olsen}}, \ and\ \bibinfo {author}
  {\bibfnamefont {D.}~\bibnamefont {Sundholm}},\ }\bibfield  {title} {\enquote
  {\bibinfo {title} {Stabilization energies of charged multiexciton complexes
  calculated at configuration interaction level},}\ }\href {\doibase
  10.1016/S1386-9477(03)00146-2} {\bibfield  {journal} {\bibinfo  {journal}
  {Physica E: Low-Dimensional Systems and Nanostructures}\ }\textbf {\bibinfo
  {volume} {18}},\ \bibinfo {pages} {436--442} (\bibinfo {year}
  {2003}{\natexlab{a}})}\BibitemShut {NoStop}%
\bibitem [{\citenamefont {Corni}\ \emph
  {et~al.}(2003{\natexlab{b}})\citenamefont {Corni}, \citenamefont
  {Brask\'{e}n}, \citenamefont {Lindberg}, \citenamefont {Olsen},\ and\
  \citenamefont {Sundholm}}]{Corni2003853141}%
  \BibitemOpen
  \bibfield  {author} {\bibinfo {author} {\bibfnamefont {S.}~\bibnamefont
  {Corni}}, \bibinfo {author} {\bibfnamefont {M.}~\bibnamefont {Brask\'{e}n}},
  \bibinfo {author} {\bibfnamefont {M.}~\bibnamefont {Lindberg}}, \bibinfo
  {author} {\bibfnamefont {J.}~\bibnamefont {Olsen}}, \ and\ \bibinfo {author}
  {\bibfnamefont {D.}~\bibnamefont {Sundholm}},\ }\bibfield  {title} {\enquote
  {\bibinfo {title} {Electron-hole recombination density matrices obtained from
  large configuration-interaction expansions},}\ }\href {\doibase
  10.1103/PhysRevB.67.085314} {\bibfield  {journal} {\bibinfo  {journal}
  {Physical Review B - Condensed Matter and Materials Physics}\ }\textbf
  {\bibinfo {volume} {67}},\ \bibinfo {pages} {853141--853147} (\bibinfo {year}
  {2003}{\natexlab{b}})}\BibitemShut {NoStop}%
\bibitem [{\citenamefont {Corni}\ \emph
  {et~al.}(2003{\natexlab{c}})\citenamefont {Corni}, \citenamefont
  {Brask\'{e}n}, \citenamefont {Lindberg}, \citenamefont {Olsen},\ and\
  \citenamefont {Sundholm}}]{Corni2003453131}%
  \BibitemOpen
  \bibfield  {author} {\bibinfo {author} {\bibfnamefont {S.}~\bibnamefont
  {Corni}}, \bibinfo {author} {\bibfnamefont {M.}~\bibnamefont {Brask\'{e}n}},
  \bibinfo {author} {\bibfnamefont {M.}~\bibnamefont {Lindberg}}, \bibinfo
  {author} {\bibfnamefont {J.}~\bibnamefont {Olsen}}, \ and\ \bibinfo {author}
  {\bibfnamefont {D.}~\bibnamefont {Sundholm}},\ }\bibfield  {title} {\enquote
  {\bibinfo {title} {Size dependence of the electron-hole recombination rates
  in semiconductor quantum dots},}\ }\href {\doibase
  10.1103/PhysRevB.67.045313} {\bibfield  {journal} {\bibinfo  {journal}
  {Physical Review B - Condensed Matter and Materials Physics}\ }\textbf
  {\bibinfo {volume} {67}},\ \bibinfo {pages} {453131--453139} (\bibinfo {year}
  {2003}{\natexlab{c}})}\BibitemShut {NoStop}%
\bibitem [{\citenamefont {V\"{a}nsk\"{a}}\ \emph {et~al.}(2006)\citenamefont
  {V\"{a}nsk\"{a}}, \citenamefont {Lindberg}, \citenamefont {Olsen},\ and\
  \citenamefont {Sundholm}}]{Vanska20064035}%
  \BibitemOpen
  \bibfield  {author} {\bibinfo {author} {\bibfnamefont {T.}~\bibnamefont
  {V\"{a}nsk\"{a}}}, \bibinfo {author} {\bibfnamefont {M.}~\bibnamefont
  {Lindberg}}, \bibinfo {author} {\bibfnamefont {J.}~\bibnamefont {Olsen}}, \
  and\ \bibinfo {author} {\bibfnamefont {D.}~\bibnamefont {Sundholm}},\
  }\bibfield  {title} {\enquote {\bibinfo {title} {Computational methods for
  studies of multiexciton complexes},}\ }\href {\doibase
  10.1002/pssb.200642169} {\bibfield  {journal} {\bibinfo  {journal} {Physica
  Status Solidi (B) Basic Research}\ }\textbf {\bibinfo {volume} {243}},\
  \bibinfo {pages} {4035--4045} (\bibinfo {year} {2006})}\BibitemShut {NoStop}%
\bibitem [{\citenamefont {V\"{a}nsk\"{a}}\ and\ \citenamefont
  {Sundholm}(2010)}]{Vanska2010}%
  \BibitemOpen
  \bibfield  {author} {\bibinfo {author} {\bibfnamefont {T.}~\bibnamefont
  {V\"{a}nsk\"{a}}}\ and\ \bibinfo {author} {\bibfnamefont {D.}~\bibnamefont
  {Sundholm}},\ }\bibfield  {title} {\enquote {\bibinfo {title} {Interpretation
  of the photoluminescence spectrum of double quantum rings},}\ }\href
  {\doibase 10.1103/PhysRevB.82.085306} {\bibfield  {journal} {\bibinfo
  {journal} {Physical Review B - Condensed Matter and Materials Physics}\
  }\textbf {\bibinfo {volume} {82}} (\bibinfo {year} {2010}),\
  10.1103/PhysRevB.82.085306}\BibitemShut {NoStop}%
\bibitem [{\citenamefont {Sundholm}\ and\ \citenamefont
  {V\"{a}nsk\"{a}}(2012)}]{Sundholm201296}%
  \BibitemOpen
  \bibfield  {author} {\bibinfo {author} {\bibfnamefont {D.}~\bibnamefont
  {Sundholm}}\ and\ \bibinfo {author} {\bibfnamefont {T.}~\bibnamefont
  {V\"{a}nsk\"{a}}},\ }\bibfield  {title} {\enquote {\bibinfo {title}
  {Computational methods for studies of semiconductor quantum dots and
  rings},}\ }\href {\doibase 10.1039/C2PC90004A} {\bibfield  {journal}
  {\bibinfo  {journal} {Annual Reports on the Progress of Chemistry - Section
  C}\ }\textbf {\bibinfo {volume} {108}},\ \bibinfo {pages} {96--125} (\bibinfo
  {year} {2012})}\BibitemShut {NoStop}%
\bibitem [{\citenamefont {Elward}\ and\ \citenamefont
  {Chakraborty}()}]{elward_atomistic}%
  \BibitemOpen
  \bibfield  {author} {\bibinfo {author} {\bibfnamefont {Jennifer~M.}\
  \bibnamefont {Elward}}\ and\ \bibinfo {author} {\bibfnamefont {Arindam}\
  \bibnamefont {Chakraborty}},\ }\bibfield  {title} {\enquote {\bibinfo {title}
  {Atomistic pseudopotential calculation on cdse quantum dots using
  electron-hole explicitly correlated hartree-fock method},}\ }\href@noop {}
  {\bibinfo  {journal} {(to be submitted)}\ }\BibitemShut {NoStop}%
\bibitem [{\citenamefont {Halonen}\ \emph {et~al.}(1992)\citenamefont
  {Halonen}, \citenamefont {Chakraborty},\ and\ \citenamefont
  {Pietil\"{a}inen}}]{RefWorks:44}%
  \BibitemOpen
\bibfield  {journal} {  }\bibfield  {author} {\bibinfo {author} {\bibfnamefont
  {V.}~\bibnamefont {Halonen}}, \bibinfo {author} {\bibfnamefont
  {T.}~\bibnamefont {Chakraborty}}, \ and\ \bibinfo {author} {\bibfnamefont
  {P.}~\bibnamefont {Pietil\"{a}inen}},\ }\bibfield  {title} {\enquote
  {\bibinfo {title} {Excitons in a parabolic quantum dot in magnetic fields},}\
  }\href {\doibase 10.1103/PhysRevB.45.5980} {\bibfield  {journal} {\bibinfo
  {journal} {Physical Review B}\ }\textbf {\bibinfo {volume} {45}},\ \bibinfo
  {pages} {5980--5985} (\bibinfo {year} {1992})}\BibitemShut {NoStop}%
\bibitem [{\citenamefont {El-Said}(1994)}]{RefWorks:43}%
  \BibitemOpen
  \bibfield  {author} {\bibinfo {author} {\bibfnamefont {Mohammad}\
  \bibnamefont {El-Said}},\ }\bibfield  {title} {\enquote {\bibinfo {title}
  {Ground-state energy of an exciton in a parabolic quantum dot},}\ }\href
  {\doibase 10.1088/0268-1242/9/3/006} {\bibfield  {journal} {\bibinfo
  {journal} {Semiconductor Science and Technology}\ }\textbf {\bibinfo {volume}
  {9}},\ \bibinfo {pages} {272--274} (\bibinfo {year} {1994})}\BibitemShut
  {NoStop}%
\bibitem [{\citenamefont {Jaziri}\ and\ \citenamefont
  {Bennaceur}(1994)}]{RefWorks:45}%
  \BibitemOpen
  \bibfield  {author} {\bibinfo {author} {\bibfnamefont {S.}~\bibnamefont
  {Jaziri}}\ and\ \bibinfo {author} {\bibfnamefont {R.}~\bibnamefont
  {Bennaceur}},\ }\bibfield  {title} {\enquote {\bibinfo {title} {Excitons in
  parabolic quantum dots in electric and magnetic fields},}\ }\href {\doibase
  10.1088/0268-1242/9/10/003} {\bibfield  {journal} {\bibinfo  {journal}
  {Semiconductor Science and Technology}\ }\textbf {\bibinfo {volume} {9}},\
  \bibinfo {pages} {1775--1780} (\bibinfo {year} {1994})}\BibitemShut {NoStop}%
\bibitem [{\citenamefont {Lamouche}\ and\ \citenamefont
  {Fishman}(1998)}]{RefWorks:46}%
  \BibitemOpen
  \bibfield  {author} {\bibinfo {author} {\bibfnamefont {G.}~\bibnamefont
  {Lamouche}}\ and\ \bibinfo {author} {\bibfnamefont {G.}~\bibnamefont
  {Fishman}},\ }\bibfield  {title} {\enquote {\bibinfo {title} {Two interacting
  electrons in a three-dimensional parabolic quantum dot: A simple solution},}\
  }\href {\doibase 10.1088/0953-8984/10/35/018} {\bibfield  {journal} {\bibinfo
   {journal} {Journal of Physics Condensed Matter}\ }\textbf {\bibinfo {volume}
  {10}},\ \bibinfo {pages} {7857--7867} (\bibinfo {year} {1998})}\BibitemShut
  {NoStop}%
\bibitem [{\citenamefont {Xie}\ and\ \citenamefont {Gu}(2003)}]{RefWorks:49}%
  \BibitemOpen
  \bibfield  {author} {\bibinfo {author} {\bibfnamefont {W.}~\bibnamefont
  {Xie}}\ and\ \bibinfo {author} {\bibfnamefont {J.}~\bibnamefont {Gu}},\
  }\bibfield  {title} {\enquote {\bibinfo {title} {Exciton bound to a neutral
  donor in parabolic quantum dots},}\ }\href {\doibase
  10.1016/S0375-9601(03)00651-0} {\bibfield  {journal} {\bibinfo  {journal}
  {Physics Letters, Section A: General, Atomic and Solid State Physics}\
  }\textbf {\bibinfo {volume} {312}},\ \bibinfo {pages} {385--390} (\bibinfo
  {year} {2003})}\BibitemShut {NoStop}%
\bibitem [{\citenamefont {Xie}(2005)}]{RefWorks:48}%
  \BibitemOpen
  \bibfield  {author} {\bibinfo {author} {\bibfnamefont {W.}~\bibnamefont
  {Xie}},\ }\bibfield  {title} {\enquote {\bibinfo {title} {Exciton states
  trapped by a parabolic quantum dot},}\ }\href {\doibase
  10.1016/j.physb.2004.12.035} {\bibfield  {journal} {\bibinfo  {journal}
  {Physica B: Condensed Matter}\ }\textbf {\bibinfo {volume} {358}},\ \bibinfo
  {pages} {109--113} (\bibinfo {year} {2005})}\BibitemShut {NoStop}%
\bibitem [{\citenamefont {Xie}(2009)}]{RefWorks:47}%
  \BibitemOpen
  \bibfield  {author} {\bibinfo {author} {\bibfnamefont {W.}~\bibnamefont
  {Xie}},\ }\bibfield  {title} {\enquote {\bibinfo {title} {Effect of an
  electric field and nonlinear optical rectification of confined excitons in
  quantum dots},}\ }\href {\doibase 10.1002/pssb.200945265} {\bibfield
  {journal} {\bibinfo  {journal} {Physica Status Solidi (B) Basic Research}\
  }\textbf {\bibinfo {volume} {246}},\ \bibinfo {pages} {2257--2262} (\bibinfo
  {year} {2009})}\BibitemShut {NoStop}%
\bibitem [{\citenamefont {Karimi}\ and\ \citenamefont
  {Rezaei}(2011)}]{Karimi20114423}%
  \BibitemOpen
  \bibfield  {author} {\bibinfo {author} {\bibfnamefont {M.J.}\ \bibnamefont
  {Karimi}}\ and\ \bibinfo {author} {\bibfnamefont {G.}~\bibnamefont
  {Rezaei}},\ }\bibfield  {title} {\enquote {\bibinfo {title} {Effects of
  external electric and magnetic fields on the linear and nonlinear
  intersubband optical properties of finite semi-parabolic quantum dots},}\
  }\href {\doibase 10.1016/j.physb.2011.08.105} {\bibfield  {journal} {\bibinfo
   {journal} {Physica B: Condensed Matter}\ }\textbf {\bibinfo {volume}
  {406}},\ \bibinfo {pages} {4423 -- 4428} (\bibinfo {year}
  {2011})}\BibitemShut {NoStop}%
\bibitem [{\citenamefont {Nammas}\ \emph {et~al.}(2011)\citenamefont {Nammas},
  \citenamefont {Sandouqa}, \citenamefont {Ghassib},\ and\ \citenamefont
  {Al-Sugheir}}]{Nammas20114671}%
  \BibitemOpen
  \bibfield  {author} {\bibinfo {author} {\bibfnamefont {F.S.}\ \bibnamefont
  {Nammas}}, \bibinfo {author} {\bibfnamefont {A.S.}\ \bibnamefont {Sandouqa}},
  \bibinfo {author} {\bibfnamefont {H.B.}\ \bibnamefont {Ghassib}}, \ and\
  \bibinfo {author} {\bibfnamefont {M.K.}\ \bibnamefont {Al-Sugheir}},\
  }\bibfield  {title} {\enquote {\bibinfo {title} {Thermodynamic properties of
  two-dimensional few-electrons quantum dot using the static fluctuation
  approximation (sfa)},}\ }\href {\doibase 10.1016/j.physb.2011.09.058}
  {\bibfield  {journal} {\bibinfo  {journal} {Physica B: Condensed Matter}\
  }\textbf {\bibinfo {volume} {406}},\ \bibinfo {pages} {4671 -- 4677}
  (\bibinfo {year} {2011})}\BibitemShut {NoStop}%
\bibitem [{\citenamefont {Rezaei}\ \emph {et~al.}(2011)\citenamefont {Rezaei},
  \citenamefont {Vaseghi},\ and\ \citenamefont {Sadri}}]{Rezaei20114596}%
  \BibitemOpen
  \bibfield  {author} {\bibinfo {author} {\bibfnamefont {G.}~\bibnamefont
  {Rezaei}}, \bibinfo {author} {\bibfnamefont {B.}~\bibnamefont {Vaseghi}}, \
  and\ \bibinfo {author} {\bibfnamefont {M.}~\bibnamefont {Sadri}},\ }\bibfield
   {title} {\enquote {\bibinfo {title} {External electric field effect on the
  optical rectification coefficient of an exciton in a spherical parabolic
  quantum dot},}\ }\href {\doibase 10.1016/j.physb.2011.09.032} {\bibfield
  {journal} {\bibinfo  {journal} {Physica B: Condensed Matter}\ }\textbf
  {\bibinfo {volume} {406}},\ \bibinfo {pages} {4596 -- 4599} (\bibinfo {year}
  {2011})}\BibitemShut {NoStop}%
\bibitem [{\citenamefont {Elward}\ \emph
  {et~al.}(2012{\natexlab{c}})\citenamefont {Elward}, \citenamefont {Hoja},\
  and\ \citenamefont {Chakraborty}}]{PhysRevA.86.062504}%
  \BibitemOpen
  \bibfield  {author} {\bibinfo {author} {\bibfnamefont {Jennifer~M.}\
  \bibnamefont {Elward}}, \bibinfo {author} {\bibfnamefont {Johannes}\
  \bibnamefont {Hoja}}, \ and\ \bibinfo {author} {\bibfnamefont {Arindam}\
  \bibnamefont {Chakraborty}},\ }\bibfield  {title} {\enquote {\bibinfo {title}
  {Variational solution of the congruently transformed hamiltonian for
  many-electron systems using a full-configuration-interaction calculation},}\
  }\href {\doibase 10.1103/PhysRevA.86.062504} {\bibfield  {journal} {\bibinfo
  {journal} {Phys. Rev. A}\ }\textbf {\bibinfo {volume} {86}},\ \bibinfo
  {pages} {062504} (\bibinfo {year} {2012}{\natexlab{c}})}\BibitemShut
  {NoStop}%
\bibitem [{\citenamefont {Bayne}\ \emph {et~al.}()\citenamefont {Bayne},
  \citenamefont {Drogo},\ and\ \citenamefont {Chakraborty}}]{bayne_pios}%
  \BibitemOpen
  \bibfield  {author} {\bibinfo {author} {\bibfnamefont {Mike}\ \bibnamefont
  {Bayne}}, \bibinfo {author} {\bibfnamefont {John}\ \bibnamefont {Drogo}}, \
  and\ \bibinfo {author} {\bibfnamefont {Arindam}\ \bibnamefont
  {Chakraborty}},\ }\bibfield  {title} {\enquote {\bibinfo {title}
  {Infinite-order diagrammatic summation approach to explicitly correlated
  congruent transformed hamiltonian},}\ }\href@noop {} {\bibinfo  {journal}
  {(to be submitted)}\ }\BibitemShut {NoStop}%
\bibitem [{\citenamefont {Swalina}\ \emph {et~al.}(2006)\citenamefont
  {Swalina}, \citenamefont {Pak}, \citenamefont {Chakraborty},\ and\
  \citenamefont {Hammes-Schiffer}}]{doi:10.1021/jp0634297}%
  \BibitemOpen
\bibfield  {journal} {  }\bibfield  {author} {\bibinfo {author} {\bibfnamefont
  {Chet}\ \bibnamefont {Swalina}}, \bibinfo {author} {\bibfnamefont
  {Michael~V.}\ \bibnamefont {Pak}}, \bibinfo {author} {\bibfnamefont
  {Arindam}\ \bibnamefont {Chakraborty}}, \ and\ \bibinfo {author}
  {\bibfnamefont {Sharon}\ \bibnamefont {Hammes-Schiffer}},\ }\bibfield
  {title} {\enquote {\bibinfo {title} {Explicit dynamical electron-proton
  correlation in the nuclear-electronic orbital framework},}\ }\href {\doibase
  10.1021/jp0634297} {\bibfield  {journal} {\bibinfo  {journal} {The Journal of
  Physical Chemistry A}\ }\textbf {\bibinfo {volume} {110}},\ \bibinfo {pages}
  {9983--9987} (\bibinfo {year} {2006})}\BibitemShut {NoStop}%
\bibitem [{\citenamefont {Jasieniak}\ and\ \citenamefont
  {Mulvaney}(2007)}]{jasieniak2007cd}%
  \BibitemOpen
  \bibfield  {author} {\bibinfo {author} {\bibfnamefont {Jacek}\ \bibnamefont
  {Jasieniak}}\ and\ \bibinfo {author} {\bibfnamefont {Paul}\ \bibnamefont
  {Mulvaney}},\ }\bibfield  {title} {\enquote {\bibinfo {title} {From cd-rich
  to se-rich-the manipulation of cdse nanocrystal surface stoichiometry},}\
  }\href {\doibase 10.1021/ja066205a} {\bibfield  {journal} {\bibinfo
  {journal} {Journal of the American Chemical Society}\ }\textbf {\bibinfo
  {volume} {129}},\ \bibinfo {pages} {2841--2848} (\bibinfo {year}
  {2007})}\BibitemShut {NoStop}%
\bibitem [{\citenamefont {Luther}\ and\ \citenamefont
  {Pietryga}(2013)}]{luther2013stoichiometry}%
  \BibitemOpen
  \bibfield  {author} {\bibinfo {author} {\bibfnamefont {Joseph~M}\
  \bibnamefont {Luther}}\ and\ \bibinfo {author} {\bibfnamefont {Jeffrey~M}\
  \bibnamefont {Pietryga}},\ }\bibfield  {title} {\enquote {\bibinfo {title}
  {Stoichiometry control in quantum dots: A viable analog to impurity doping of
  bulk materials},}\ }\href {\doibase 10.1021/nn401100n} {\bibfield  {journal}
  {\bibinfo  {journal} {ACS nano}\ }\textbf {\bibinfo {volume} {7}},\ \bibinfo
  {pages} {1845--1849} (\bibinfo {year} {2013})}\BibitemShut {NoStop}%
\end{thebibliography}%
